\documentclass{aa}
\usepackage{graphicx}

\begin{document}

\newcommand{\fe}{\ion{Fe}{ii}}
\newcommand{\h}{H$_2$}
\newcommand{\ci}{\ion{C}{i}}
\newcommand{\oi}{\ion{O}{i}}
\newcommand{\s}{\ion{S}{ii}}
\newcommand{\n}{\ion{N}{i}}
\newcommand{\he}{\ion{He}{i}}
\newcommand{\nii}{\ion{N}{ii}}

\title{Molecular line emission in HH54: a coherent view from near to far infrared
\thanks{Based on observations collected at the European Southern Observatory (La Silla and 
Paranal), Chile (65.I-0150, 70.C-0138, 74.C-0235) and observations with ISO,pa an ESA project with 
instruments funded by ESA Members States (especially the PI countries: France, Germany, the
Netherlands, and the United Kingdom) and with the participation of ISAS and NASA.}}
\author{T. Giannini$^1$, C. M$^{\rm c}$Coey$^2$, B. Nisini$^1$, S.Cabrit$^3$, 
A.Caratti o Garatti$^4$, L.Calzoletti$^{1,5}$, D.R. Flower$^6$}
\offprints{Teresa Giannini, email:giannini@oa-roma.inaf.it}
\institute{{$^1$ INAF-Osservatorio Astronomico di Roma, via Frascati 33, I-00040 Monte Porzio Catone, Italy\\
$^2$ Department of Physics, University of Waterloo 200, University Avenue W.,Ontario, N2L 3G1, Canada\\
$^3$ LERMA, Observatoire de Paris, UMR 8112, France\\
$^4$ Th\"{u}ringer Landessternwarte, Sternwarte 5, 07778 Tautenburg, Germany\\
$^5$ Universit\`a di Cagliari, Via Universit\`a 40, 09124 Cagliari, Italy\\
$^6$ Physics Department, Durham University, Durham, DH1 3LE, UK}\\
\email{giannini,nisini,calzol@oa-roma.inaf.it;cmccoey@astro.uwaterloo.ca;\\
sylvie.cabrit@obspm.fr;caratti@tls-tautenburg.de;david.flower@durham.ac.uk}}
%
%
\date{Received date; Accepted date}
%
%
%
\titlerunning{Molecular emission in HH54}
\authorrunning{T.Giannini et al.}

\abstract{
{\it Aims.} We present a detailed study of the infrared line emission (1-200$\mu$m) in the Herbig-Haro object 
HH54. Our database comprises: high- (R$\sim$ 9000) and low- (R$\sim$600) resolution spectroscopic data in the near-infrared 
band (1-2.5$\mu$m); mid-infrared  spectrophotometric images (5-12$\mu$m); and, far-IR (45-200$\mu$m, R$\sim$200) 
spectra acquired with the ISO satellite. As a result, we provide the detection of and the absolute fluxes for more than 60 molecular features 
(mainly from H$_2$ in the near- and mid-infrared and from H$_2$O, CO and OH in the far-infrared) and 23 ionic lines.\\
{\it Methods.} The H$_2$ lines, coming from levels from $v$=0 to $v$=4 have been interpreted in the context of a state-of-art shock
code, whose output parameters are adopted
as input to a Large Velocity Gradient computation in order to interpret the FIR emission of CO, H$_2$O 
and OH.\\
{\it Results.} The H$_2$ emission can be interpreted as originating in either steady-state J-type shocks or in quasi-steady J-type shocks with magnetic 
precursor.  However, our multi-species analysis shows that only a model of a J-type shock with magnetic precursor 
(v$_{\mathrm{shock}}$=18 km\,s$^{-1}$, n$_{\mathrm{H}}$=10$^{4}$ cm$^{-3}$, 
B=100\,$\mu$G, age=400 yr) can account for both the observed \h \, emission \emph{and} the CO and H$_2$O lines. Such a model predicts a H$_2$O abundance of
$\sim$ 7 10$^{-5}$, in agreement with estimations from other shock models for outflows associated with low mass protostars. 
We can exclude the possibility that the observed atomic lines arise in the same shock as the molecular lines, and give arguments in favour 
of the presence of a further high-velocity,
fully dissociative shock component in the region. Finally, in view of the forthcoming spectroscopic facilities on board of the Herschel satellite, 
we provide predictions for H$_2$O lines considered to be the most suitable for diagnostic purposes. 
} 
\maketitle  
%
%


\section{Introduction}
It is now commonly accepted that the earliest stages of star formation are accompanied by powerful bipolar jets, which act to remove excess angular 
momentum from in-falling gas (e.g. Bachiller, 1996).  The jets are generally highly collimated and often terminate in a curve-shaped structure (bow-shock); 
however, knots of shocked emission, which may display a bow morphology, are also observed along the jet (e.g. Reipurth \& Bally, 2001). 
The thermal energy associated with a shock is radiated away through the emission of lines from atomic, ionic and molecular species.   
Gas at temperatures of thousands of degrees Kelvin cools principally through ro-vibrational lines of molecular
hydrogen, in the near-infrared, but also through forbidden transitions of abundant atomic and ionic species, such as [\fe],[\s],[\ci] and [\n].  Colder gas components, at hundreds 
of degrees Kelvin, cool via mid- and far-infrared molecular lines, particularly rotational transitions of H$_2$ (at $\lambda$ $\le$ 28 $\mu$m) and
lines of other molecular species, such as H$_2$O, CO and OH (in the far-infrared). The relative contribution to the 
gas cooling depends strongly upon the physical conditions within the shock.
In particular, the shock velocity and the conditions in the pre-shock gas (the degree of ionization, the strength of the 
magnetic field and the density) determine the structure of the shock wave (Draine, 1980), which is usually classified as 
'jump' (J-type), 'J-type with magnetic precursor' or 'continuous' (C-type) (e.g., Hollenbach \& McKee 1989). In 
general, J-type shocks produce a narrower profile 
and a higher peak temperature than C-type shocks while J-type shocks with magnetic precursor can be considered to be intermediate.  J-type shocks 
can heat the gas up to temperatures of 10$^5$ K and can travel at up to hundreds of km\,s$^{-1}$, while C-type shock velocities do not exceed 
$\approx$ 80 km s$^{-1}$ (Le Bourlot et al. 2002, Flower et al., 2003).  Classically, calculations of shock structure have been performed under 
the assumption of steady state.  However, if this assumption is relaxed it has been shown that, in the presence of a sufficiently strong magnetic 
field, an initially J-type discontinuity may evolve in time into a C-type shock wave (Smith \& MacLow, 1997, Chi\`eze, Pineau des For\^{e}ts \& Flower, 1998).  
The intermediary stage of the evolution is described by a (quasi-steady) J-type shock wave with magnetic precursor and can be considered as a J-type 
discontinuity introduced into a C-type shock profile at the point in the ion fluid flow time, $t_\mathrm{i} = \int 1 / v_\mathrm{i} dz$, that may be
identified with the age of the shock (Lesaffre et al., 2004a).

Although 2D and 3D hydrodynamical and MHD simulations of jet-driven molecular shocks have progressed
enormously over the last years, including some time-dependent H$_2$ chemistry,
H ionization, and cooling (Downes \& Cabrit, 2003; Smith \& Rosen, 2003),
the effect of ion-neutral decoupling in magnetic precursors and the
resulting complex shock chemistry have so far only been simulated in 1D
(Chieze et al., 1998; Lesaffre et al., 2004b). Lesaffre et al. (2004a) showed
that truncated steady-state multi-fluid models provide a
useful and efficient approximation to time-dependent simulations, while
allowing a more detailed treatment of the chemistry and hence of the degree of ionization.  We shall
thus adopt this approach here.
\\
Since different shock types give rise to very different excitation conditions, the 
interpretation of multiple species observed over a wide range in wavelength provides strong constraints for a shock model.  Multi--wavelength 
studies have already proved useful: for example, Smith, Froebrich \& Eisl\"{o}ffel (2003) employed an analysis of the near- to far-infrared 
molecular emission observed towards Cepheus E in order to determine both the shock type and geometry.  The main limitation of the 
multi--wavelength/multi--species approach arises from the poor sensitivity and, more importantly, the poor spatial resolution achieved in the 
mid- and far-infrared by space-borne instrumentation, which often prevents the resolution of structures along jets and in bow-shocks. In the near 
future, however, we expect that these limitations will be overcome by the instruments aboard the Herschel satellite, whose performance in terms
of sensitivity and of spatial and spectral resolution will be closer to those attainable with ground-based observations at shorter wavelengths. 

The aims of the present work are threefold.  Firstly, we wish to test the ability of a current shock model to self-consistently reproduce emission 
from several different species observed over a wide range of wavelengths. Secondly, we will investigate which combinations of these observations 
prove most useful in determining shock conditions, particularly shock type.  Finally, from this analysis, we intend to identify important diagnostic 
molecular lines, mainly of H$_2$O, falling in the wavelength range observable with Herschel. 
As a test case, we have applied our analysis to the Herbig-Haro object HH54. This is located at the northern edge of 
the nearby star-forming region ChaII (d$\approx$ 200 pc, Hughes \& Hartigan, 1992). The object, consisting of a complex of 
several arcsecond-scale bright knots, was originally discovered by Schwartz (1977), then spectroscopically observed in the optical 
by Schwartz \& Dopita (1980) and Graham \& Hartigan (1988), who 
found a blue-shifted gas with a radial velocity that becomes increasingly blue in a long 'streamer' from the north to the south and that may be
 a jet responsible for the
excitation of the HH54 complex. Knee (1992, hereafter K92) associates HH54 with a monopolar blue-shifted CO
outflow, whose driving source remains unclear after a proper motion analysis of the infrared knots (Caratti o Garatti, et al. 2006). 
Molecular hydrogen emission was first reported by Sandell et al. (1987), while
the near-infrared spectrum was obtained by Gredel (1994, hereafter G94). In the mid-infrared, observations with ISO/CAM 
(Cabrit, et al. 1999), ISO/SWS (Neufeld, Melnick \& Harwit, 1998), and quite recently, with Spitzer/IRS (Neufeld et al., 2006) have shown the
H$_2$ ortho-to-para ratio to lie below the equilibrium value expected at the observed gas temperature.
Far-infrared emission, detected with ISO/LWS has been reported by Nisini et al. (1996) 
and discussed by Liseau et al. (1996) in the framework 
of planar, steady-state C-shocks.  Conversely, Wilgenbus et al. (2000), who compared the predictions of steady-state shock models to observations 
of pure rotational (Neufeld et al., 1998) and ro-vibrational (G94) \h \, lines, concluded that no single planar C- or J-type shock
is able to account for the observed emission in both pure rotational and ro-vibrational lines.
In the present work, we will check whether or not this result can be confirmed on the basis of a large observational data-base.  

The structure of the paper is the following: in Sect.2 the observations used for our analysis are presented, while in Sect. 
3 the adopted shock model is described and applied to the data. Our conclusions are then summarized in Sect.4.

\section{Observations and Results}
Our database consists of data collected with the ESO facilities and retrieved from  the ISO archive 
\footnote {available at http://www.iso.vilspa.esa.es/ida/index.html}. The relevant information on the observational 
settings is summarized as a log in Table \ref{tab:journal}.

\begin{table*}
\caption{\label{tab:journal}Journal of observations.}  \small
\begin{center}
\begin{tabular}{cccccc}
\multicolumn{6}{c} {Imaging}\\
\hline
Telescope/Instrument& Date         & Filter             & FoV                                & Pixel Scale              & Exposure Time  \\
                    &              &                    & ($^{\prime})\times(^{\prime}$)     &($^{\prime\prime}$/pxl)   &          (s)   \\
\hline
NTT/SofI            & 6 Jun 1999   & H$_2$2.12$\mu$m    & 4.9$\times$4.9                     & 0.29                     & 300            \\
ISO/ISOCAM          & 20 Oct 1997  & CVF 5.0-16.8$\mu$m & 3$\times$3                         & 6                        & 3600           \\
\hline
\end{tabular}
\end{center}
\begin{center}
\begin{tabular}{ccccccc}
\multicolumn{7}{c} {Spectroscopy}\\
\hline
Telescope/Instrument& Date        & $\lambda$        & slit/aperture & P.A.	   & $\lambda$/$\Delta$$\lambda$     & Exposure Time \\
                    &             &   ($\mu$m)       &               & ($^{\circ}$)&				     &  (s) \\
\hline
VLT/ISAAC              & 11 Jul 2002  & 0.98-2.50    & I             & 28	   & 1600			     & 1600 \\
NTT/SofI               & 6 Jun 1999   & 1.53-2.52    & II            & 140	   & 600			     & 2400 \\
VLT/ISAAC              & 31 Dec 2004  & 2.06-2.18    & I/III         & 28/160	   & 8900			     & 1800 \\
ISO/LWS                & 17 Feb 1996  & 43-197       & $\theta$$\sim$ 80$^{\prime\prime}$  &-       & 200 & 2740 \\
ISO/LWS -FP            & 23 Sep 1997  & 63.13-63.23  & $\theta$$\sim$ 80$^{\prime\prime}$  &-      & 8500 & 6000 \\
\hline

\end{tabular}
\end{center}
\end{table*}

\subsection{Ground-based observations}
\subsubsection{Imaging}

\begin{figure}
\centering
\includegraphics[width=8.5cm]{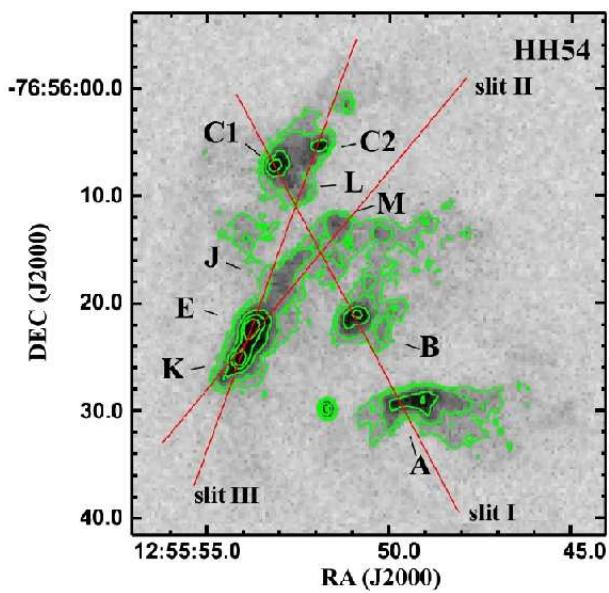}
\caption{\label{fig:H2} H$_2$ 2.12\,$\mu$m (continuum-subtracted) SofI image of HH54.
The knots of emission are labelled according to the nomenclature of G94 and
Sandell et al. (1987). The contours are 3,5,10,15 and 20$\times$ the standard deviation
to the mean background (5$\times$10$^{-15}$ erg s$^{-1}$ cm$^{-2}$ arcsec$^{-2}$).
The orientations of the slit used for spectroscopic observations
are depicted.  }
\end{figure}

A narrow-band image in the H$_2$ 1-0S(1) filter was obtained with SofI (Lidman, Cuby \& Vanzi, 2002)
at the New Technology Telescope (NTT, La Silla, Chile). The observation was conducted 
by nodding and jittering the telescope around the pointed position ($\alpha$(J2000)=12$^h$ 55$^m$ 50.3$^s$,
$\delta$(J2000)= -76$^{\circ}$ 56$^{\prime}$ 23$^{\prime\prime}$) in the usual ABB'A'mode.
The raw data were reduced by using
standard procedures for bad pixel removal, flat fielding and sky subtraction. Flux calibration was
derived from the observation of a photometric standard star during the night. A continuum-free
image was obtained by subtracting from the narrow band image an appropriately scaled K$_s$ image,
acquired just after the observation. The H$_2$, continuum-subtracted image is
shown in Fig.\ref{fig:H2}, where the knots of line emission are labelled according to the notation
of G94 and Sandell et al.\,(1987). The HH54 complex appears to be composed of arcsecond-scale bright knots: 
those labelled A, B and C lie along the axis of the optical ``streamer'' extending south of HH54; for brevity, we will 
refer to this axis as the ``jet axis'' in the following. In addition, two further knots (E,K) plus (at least) three more diffuse
regions are characterized by a bow-shaped morphology. Additional knots, roughly located 
along the axis and barely visible in the G94 image, have been detected (namely, J,L,M). 
The (1-0)S(1) line flux of the whole HH54 complex is 1.12$\times$10$^{-12}$ erg s$^{-1}$ cm$^{-2}$,
i.e. approximately a factor of two and five larger than the values measured by G94 and Sandell (1987), 
respectively. The brightness variability of the H$_2$
1-0 S(1) line on a time scale of about 10 years, pointed out by G94, is thus confirmed
by the present observations. As suggested by Liseau et al. (1996), such variability indicates that
HH54 is a non-equilibrium flow.

\subsubsection{Low-resolution spectroscopy}

\begin{table*}
\caption{\label{tab:fluxes} Near infrared line fluxes observed in knot B.}  \small
\begin{center}
\begin{tabular}{ccc|ccc}
\hline\\[-5pt]
Line id.                   &  $\lambda$       & $F\pm\Delta~F$$^a$                & Line id.                                    &  $\lambda$ & $F\pm\Delta~F$$^a$ \\
                                        &   ($\mu$m)       & (10$^{-15}$erg\,cm$^{-2}$\,s$^{-1}$)      &                                         &  ($\mu$m)  & (10$^{-15}$erg\,cm$^{-2}$\,s$^{-1}$)\\
\hline\\[-5pt]
\multicolumn{6}{c}{H$_2$ lines} \\
\hline\\[-5pt]
2--0 S(9)                               &     1.053    &        0.66$\pm$0.07              & 3--1 Q(7)		             &	 1.368   &   0.4 $\pm$0.1  \\  
2--0 S(7)                               &     1.064    &        0.62$\pm$0.04              & ~1--0 S(9) 		     &	 1.688	 &   0.7 $\pm$0.1  \\ 
2--0 S(6)                               &     1.073    &        0.4 $\pm$0.1               & ~1--0 S(8)$^d$		     &	 1.715	 &   0.6 $\pm$0.1  \\ 
2--0 S(5)                               &     1.085    &        0.69$\pm$0.05              & ~1--0 S(7)$^e$		     &	 1.748	 &   3.9 $\pm$0.1  \\ 
2--0 S(4)                               &     1.100    &        0.3 $\pm$0.2$^b$           & ~1--0 S(6) 		     &	 1.788	 &   1.86$\pm$0.05 \\ 
3--1 S(9)+3--1 S(10)+3--1S(11)          &     1.120-1.120-1.121&1.4 $\pm$0.5$^b$           & ~1--0 S(3) 		     &	 1.958	 &  15.4 $\pm$0.2  \\ 
3--1 S(7)                               &     1.130    &        0.8 $\pm$0.3$^b$           & ~2--1 S(4) 		     &   2.004   &   0.42$\pm$0.09 \\
2--0 S(2)                               &     1.138    &        0.6 $\pm$0.1               & ~1--0 S(2) 		     &   2.034   &   4.20$\pm$0.06 \\
3--1 S(5)                               &     1.152    &        0.9 $\pm$0.1               & ~2--1 S(3) 		     &   2.073   &   1.63$\pm$0.04 \\
2--0 S(1)                               &     1.162    &        1.2 $\pm$0.1               & ~1--0 S(1) 		     &   2.122   &   9.59$\pm$0.05 \\
3--1 S(3)                               &     1.186    &        0.8 $\pm$0.1               & ~2--1 S(2) 		     &   2.154   &   0.51$\pm$0.05 \\
2--0 S(0)+4--2 S(10)                    &     1.189-1.190  &    0.4 $\pm$0.1               & ~3--2 S(3) 		     &   2.201   &   0.3 $\pm$0.1  \\
4--2 S(9)                               &     1.196    &        0.4 $\pm$0.1               & ~1--0 S(0) 		     &   2.223   &   2.58$\pm$0.05 \\
4--2 S(8)+4--2 S(11)                    &     1.198-1.199  &    0.5 $\pm$0.2$^b$           & ~2--1 S(1) 		     &   2.248   &   1.0 $\pm$0.2  \\
4--2 S(7)                               &     1.205    &        0.3 $\pm$0.1               & ~2--1 S(0) 		     &   2.355   &   0.32$\pm$0.08 \\
3--1 S(2)                               &     1.207    &        0.4 $\pm$0.1               & ~3--2 S(1) 		     &   2.386   &   0.5 $\pm$0.1  \\
4--2 S(5)                               &     1.226    &        0.4 $\pm$0.1               & ~1--0 Q(1) 		     &   2.407   &  10.9 $\pm$0.3  \\
3--1 S(1)                               &     1.233    &        0.5 $\pm$0.1               & ~1--0 Q(2) 		     &   2.413   &   4.5 $\pm$0.4  \\
2--0 Q(1)                               &     1.238    &        0.55$\pm$0.05              &~ 1--0 Q(3) 		     &   2.424   &   9.6 $\pm$0.4  \\
2--0 Q(2)+4--2 S(4)                     &     1.242-1.242  &    0.39$\pm$0.04              & ~1--0 Q(4) 		     &   2.437   &   4.2 $\pm$0.5  \\
2--0 Q(3)$^c$                           &     1.247    &        0.56$\pm$0.04              & ~1--0 Q(5) 		     &   2.455   &   3.2 $\pm$0.5  \\
2--0 O(3)                               &     1.335    &        0.39$\pm$0.07              & ~1--0 Q(6) 		     &   2.476   &   2.2 $\pm$0.4  \\ 
3--1 Q(5)+4--2S(0)                      &     1.342-1.342 &     0.5$\pm$0.2$^b$            & ~1--0 Q(7) 		     &   2.500   &   6.2 $\pm$0.6  \\   
                                                                                           
\hline\\[-5pt]
\multicolumn{6}{c}{Ionic  lines} \\
\hline\\[-5pt]
~[{\ion{C}{i}}]\,$^1\!D_{2}-^3\!P_{1}$          &     0.983        &            0.98$\pm$0.06                  & ~[{\ion{Fe}{ii}}]\,a$^4\!D_{5/2}-a^6\!D_{5/2}$  &   1.295   &   1.4 $\pm$0.2  \\
~[{\ion{C}{i}}]\,$^1\!D_{2}-^3\!P_{2}$          &     0.985        &            2.33$\pm$0.05                  & ~[{\ion{Fe}{ii}}]\,a$^4\!D_{7/2}-a^6\!D_{7/2}$  &   1.321   &   2.9 $\pm$0.2  \\
~[{\ion{S}{ii}}]\,$^2\!P_{3/2}-^2\!D_{3/2}$     &     1.029        &            0.21$\pm$0.07                  & ~[{\ion{Fe}{ii}}]\,a$^4\!D_{5/2}-a^6\!D_{3/2}$  &   1.328   &   0.6 $\pm$0.1  \\
~[{\ion{S}{ii}}]\,$^2\!P_{3/2}-^2\!D_{5/2}$     &     1.032        &            0.30$\pm$0.07                  & ~[{\ion{Fe}{ii}}]\,a$^4\!D_{5/2}-a^4\!F_{9/2}$  &   1.534   &   1.2 $\pm$0.2  \\
~[{\ion{S}{ii}}]\,$^2\!P_{1/2}-^2\!D_{3/2}$     &     1.034        &            0.28$\pm$0.09                  & ~[{\ion{Fe}{ii}}]\,a$^4\!D_{3/2}-a^4\!F_{7/2}$  &   1.600   &   0.7 $\pm$0.2  \\
~[{\ion{S}{ii}}]\,$^2\!P_{1/2}-^2\!D_{5/2}$     &     1.037        &            0.2 $\pm$0.1$^b$               & ~[{\ion{Fe}{ii}}]\,a$^4\!D_{7/2}-a^4\!F_{9/2}$  &   1.644   &   8.2 $\pm$0.1  \\
~ HeI\,2$^3\!S-2^3\!P^0$                        &     1.083        &            0.53$\pm$0.05                  & ~[{\ion{Fe}{ii}}]\,a$^4\!D_{1/2}-a^4\!F_{5/2}$  &   1.664   &   0.28$\pm$0.07 \\
 ~Pa$\gamma$                                    &     1.094        &            $<$ 0.8$^f$                    & ~[{\ion{Fe}{ii}}]\,a$^4\!D_{5/2}-a^4\!F_{7/2}$  &   1.678   &   0.58$\pm$0.08 \\
~[{\ion{Fe}{ii}}]\,a$^4\!D_{7/2}-a^6\!D_{9/2}$  &     1.257        &            8.7 $\pm$0.1                   & ~[{\ion{Fe}{ii}}]\,a$^4\!D_{3/2}-a^4\!F_{3/2}$  &   1.798   &   0.4 $\pm$0.1  \\
~[{\ion{Fe}{ii}}]\,a$^4\!D_{3/2}-a^6\!D_{3/2}$  &     1.279        &            1.1 $\pm$0.2                   & ~[{\ion{Fe}{ii}}]\,a$^4\!D_{7/2}-a^4\!F_{7/2}$  &   1.810   &   3.4 $\pm$0.4  \\
~Pa$\beta$                                      &     1.283        &            1.5 $\pm$0.4                   & ~Br$\gamma$                                     &   2.166   &     $<$0.2$^f$  \\   
                       
\hline\\[-5pt]
\end{tabular}
\end{center}
Notes: $^a$ The uncertainty refers to the {\it rms} of the local baseline.\\
$^b$ signal-to-noise ratio between 2 and 3.\\
$^c$ blended with ~[{\ion{Fe}{ii}}]\,a$^4\!D_{3/2}-a^6\!D_{5/2}$ ($\lambda$=1.248 $\mu$m).\\
$^d$ blended with ~[{\ion{Fe}{ii}}]\,a$^4\!D_{3/2}-a^4\!F_{5/2}$ ($\lambda$=1.711 $\mu$m).\\
$^e$ blended with ~[{\ion{Fe}{ii}}]\,a$^4\!P_{3/2}-a^4\!D_{7/2}$ ($\lambda$=1.749 $\mu$m).\\
$^f$ 3 sigma upper limit.\\
\end{table*}
\begin{table}
\caption{\label{tab:hr} Medium-resolution spectroscopy parameters.}  \small
\begin{center}
\begin{tabular}{ccccc}
\hline
Knot &  slit & v-v$_{cloud}$$^a$  &  $\Delta$v   \\  
     &       &\multicolumn{2}{c} {(km s$^{-1}$)}  \\ 
\hline
C1   &   III   & -15           & 19$\pm$5        \\  
B    &   III   & -12           & 13$\pm$5         \\ 
A    &   III   & -12           & 16$\pm$5         \\ 
C2   &   IV    & -14           & 12$\pm$8         \\ 
L    &   IV    & -13           & 20$\pm$6         \\ 
E    &   IV    & -12           & 13$\pm$7         \\ 
K    &   IV    & -9            & 21$\pm$8         \\ 
\hline
\end{tabular}
\end{center}
Note: $^a$  The estimated uncertainty is within 3 km s$^{-1}$.\\
\end{table}

\begin{figure*}
\centering
\includegraphics[width=15cm]{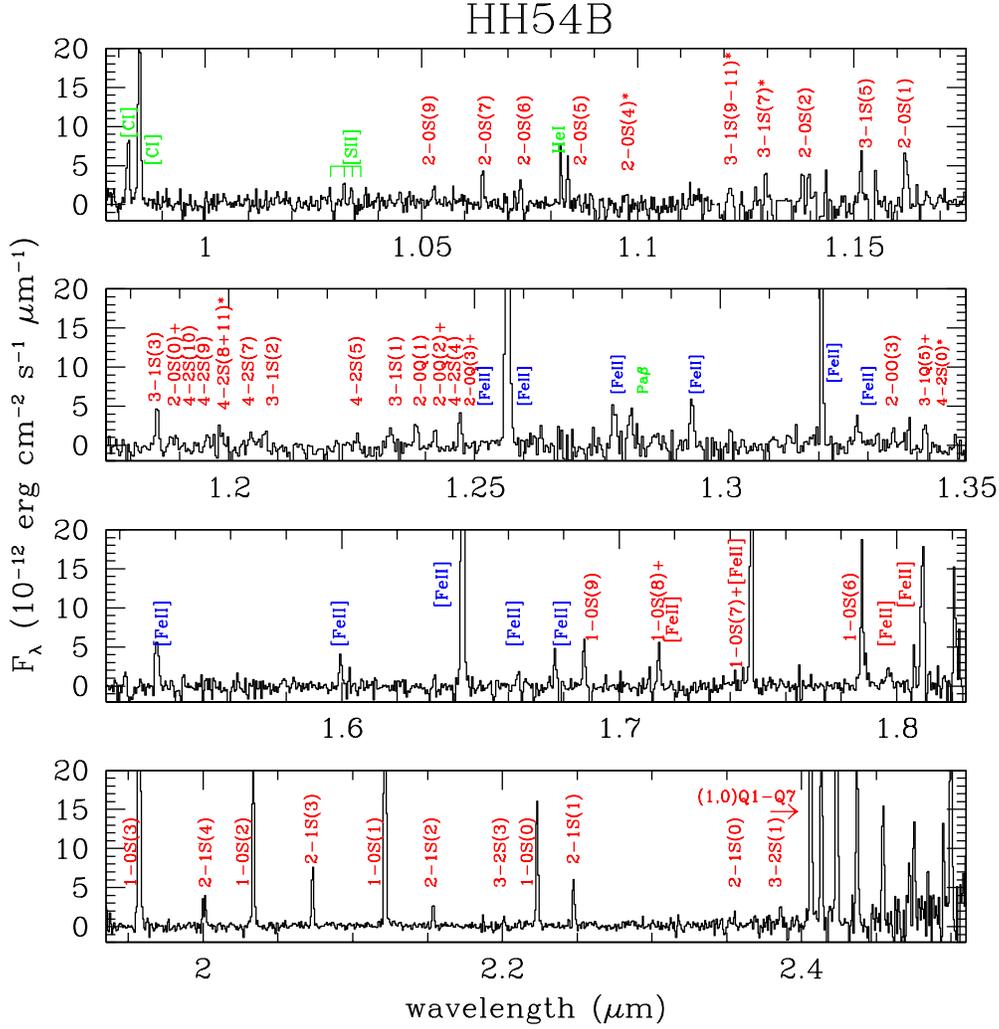}
\caption{\label{fig:spectrum}1.0-2.5 $\mu$m low resolution spectrum of HH54B. Lines marked with an asterisk
are detected with signal-to-noise ratio less than 3.}
\end{figure*}

\begin{figure*}
\centering
\includegraphics[width=18cm]{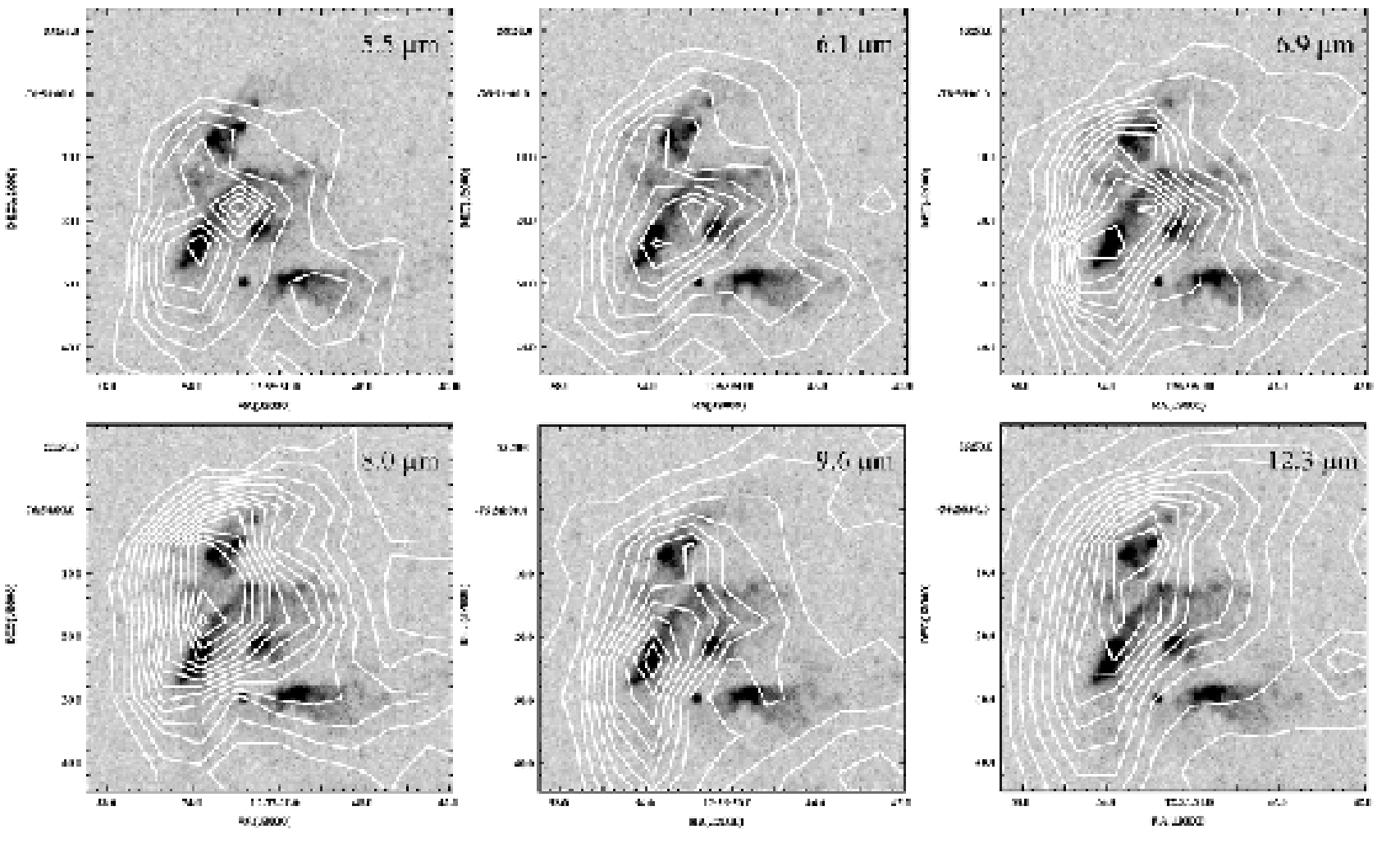}
\caption{\label{fig:isocam} ISOCAM images of HH54. Contours are in steps of 2$\sigma$ from a 3\,$\sigma$ level of
8.3 10$^{-16}$ erg s$^{-1}$ cm$^{-2}$ arcsec$^{-1}$ (at 5.5\,$\mu$m, 6.1\,$\mu$m and 6.9\,$\mu$m),
2.7 10$^{-16}$ erg s$^{-1}$ cm$^{-2}$ arcsec$^{-1}$ (at 8.0\,$\mu$m),
1.4 10$^{-15}$ erg s$^{-1}$ cm$^{-2}$ arcsec$^{-1}$  (at 9.6\,$\mu$m) and
5.5 10$^{-16}$ erg s$^{-1}$ cm$^{-2}$ arcsec$^{-1}$ (at 12.3\,$\mu$m).}
\end{figure*}

Long-slit, near infrared spectra of the HH54 region were collected with the ESO facilities
during two different runs. In 1999, we obtained the 1.55-2.50 $\mu$m spectra of the knots
along the  brightest wing of the bow (knots E,K) with SofI, while the jet axis was targetted in 2002 with ISAAC (Cuby et al., 2004)
at the Very Large Telescope (VLT, Paranal, Chile). Four grating settings  (SZ,\,J,\,SH,\,SK filters) are
required with ISAAC to cover the spectrum between  0.98-2.50 $\mu$m.
The adopted slit orientations along with the knots encompassed by the slit are depicted in Fig.\,\ref{fig:H2}.
The raw spectral images were flat-fielded, sky subtracted and corrected
for the optical distortions along both the spatial and spectral directions. Telluric features were removed
by dividing the extracted spectra by that of a blackbody-normalized telluric standard star (spectral type O9V), once corrected
for its intrinsic spectral features. Wavelength calibration was derived from the lines of a Xenon-Argon lamp, while
flux calibration was based on the photometry of the H$_2$ 1-0S(1) continuum-subtracted image, whose associated
uncertainty is within 10\%.
Ratios of lines falling in different portions of the ISAAC spectrum are affected by an uncertainty of up to 20\%, due to
the different instrumental responses in the four segments. This error was estimated by comparing the fluxes 
of lines present in overlapping parts of the spectrum.
Fig.\,\ref{fig:spectrum} and Table \ref{tab:fluxes} show the ISAAC spectrum and the line fluxes detected in knot B, 
which is among the knots richest in lines. Clearly, regions at different degrees of dissociation and ionization are encompassed
by the slit aperture: low excitation conditions are characterized by the presence of copious H$_2$ ro-vibrational lines (with
$v$ $\le$ 4 and a maximum excitation energy of $\sim$ 25000 K), while the presence of ionized gas at T $\approx$ 10000 K is evidenced 
from the detection of several forbidden lines from [\fe],[\ci],[\s], [\n] along with hydrogen and helium recombination lines.

\subsubsection{2.12\,$\mu$m medium-resolution spectroscopy}\label{sec:highres}
Medium-resolution spectra of the H$_2$ 2.1218\,$\mu$m line were obtained in January 2004 with ISAAC, along
the directions depicted in Fig.\,\ref{fig:H2} (slits I and III). The slit aperture is 0.3$^{\prime\prime}$ $\times$
120$^{\prime\prime}$, which gives a nominal resolution of about 8900, i.e. 33.7 km s$^{-1}$. The data were acquired
and reduced with the same techniques outlined in the previous Section. Wavelength calibration was performed using 
bright OH atmospheric lines (Rousselot et al., 2000), which provide an accuracy to within 3 km s$^{-1}$.
The instrumental profile in the dispersion direction, as measured from Gaussian fits to sky lines, was $\sim$ 31.7 km s$^{-1}$.
We give in Table \ref{tab:hr} the velocity parameters for all the encompassed knots.
The peak radial velocity (v-v$_{cloud}$) was computed with respect to the ambient molecular cloud, for which a velocity of
2.0 km\,s$^{-1}$ in the local standard of rest (LSR) has been adopted (K92). All the knots appear blue-shifted by about
10 km s$^{-1}$, a circumstance which, although confirming the association of HH54 with a blue-shifted outflow,
excludes the possibility that H$_2$ emission originates in the same gas that gives rise to the optical spectrum, where v$_{rad}$ ranges 
from -20 to -70 km s$^{-1}$ (Graham \& Hartigan, 1998).
The observed FWHM is larger than the instrumental width by a few km s$^{-1}$: the deconvolved values, which give
an estimate of the velocity spread ($\Delta v$), range between 13 and 20 km s$^{-1}$.  
The poor spectral resolution, coupled with the low signal-to-noise ratio at the base of the line, prevents an accurate measurement of the intrinsic FWZI, 
which could provide some information about the speed of the bow-shock (Hartigan, Raymond \& Hartmann, 1987). An average velocity of
$\approx$ 50 km\,s$^{-1}$ along the jet axis,
however, has been recently estimated from proper motion analysis  by Caratti o Garatti et al. (2006). We can consider this value
to be an upper limit to the shock velocity, given the possibility that the observed shocks travel 
in a medium already put into motion by previous ejection events: a phenomenon that is frequently observed in the 
environments of protostellar outflow (e.g. Chrysostomou et al., 2000).

\subsection{Space-borne observations}
\subsubsection{Mid-infrared spectro-imaging}
Images of HH54 in the H$_2$ pure rotational transitions from  0-0 S(2) to S(7) were obtained
with the infrared camera ISOCAM (Cesarsky et al., 1996) aboard ISO, in the guaranteed time CAMFLOWS. Spectral scans were obtained 
with the Circular Variable Filter (CVF) over a wavelength range of 16.61 to 5.08 $\mu$m with full spectral sampling, and a camera pixel size of
6$^{\prime\prime}$ yielding a total field of view of about 3$^{\prime}$ by 3$^{\prime}$. The raw data files were retrived from the ISO
archive and reprocessed locally  with the CAM Interactive Analysis (CIA; Ott et al., 2001) software.
The data reduction procedure consists of dark current and flat field correction and cosmic ray removal. Continuum was then removed under each H$_2$ line by fitting a linear baseline, and the line
flux was obtained by trapezoidal integration over 5 spectral elements, yielding individual images in each H$_2$ line. A preliminary report of this dataset was presented
by Cabrit et al. (1999).
In the reduced images, shown in Fig.\ref{fig:isocam}, the same bow-shape morphology delineated by the 1-0S(1) line is recognizable but 
the gas traced by the rotational lines appears smooth, rather than clumpy as in the 2.12\,$\mu$m image. This could be ascribed to the poor spatial resolution
at mid-infrared wavelengths, but could also reflect a different mechanism at the origin of the excitation of
the ro-vibrational and pure rotational lines. We will address this point further on in the paper (see Sec.\ref{h2}).

The extent of the regions where emission is detected at over 3$\sigma$ above the sky level, increases from the highest frequency 
line 0-0S(7) at 5.5\,$\mu$m to the 0-0S(2) line at 12.3$\mu$m, an effect which is due both to the loss of spatial resolution at increasing 
wavelength and to the progressively lower temperatures (from thousands to hundreds of Kelvin) probed by lines with lower excitation energy.
As one of our aims is to obtain a meaningful comparison between far- and near-infrared lines, we have superposed the 0-0 contours 
on the 2.12\,$\mu$m image, so as to identify the regions where the emission overlaps. The ISOCAM spatial resolution 
does not allow us to disentangle structures with an angular size of less than 6 arcsec, a dimension often larger than that of the 
near infrared knots. In the 5.5\,$\mu$m image, which better matches the morphology delineated at 2.12$\mu$m, we are able to resolve 
four peaks roughly corresponding to the three knots from north to south (knots C to A) and to the wing (knots E,K). Aiming to attribute a mid-infrared flux to each peak, we have defined four zones, each centred on the individual
peak, and containing at least 10\% of the peak emission; inside each zone the flux has been computed by using  the task {\it polyphot} in the IRAF 
package\footnote{IRAF (Image Reduction and Analysis Facility) is distributed by the National Optical Astronomy Observatories,
which are operated by AURA, Inc., cooperative agreement with the National Science Foundation.}.
The results are given in Table \ref{tab:isocam}, where each zone has been identified by indicating the corresponding
knots at 2.12\,$\mu$m and the area over which the photometry has been computed. Since fluxes of mid-infrared lines are poorly 
affected by the extinction (which is in any case low, as will be shown in Sect.3.1), their ratios directly give a precise indication 
of the local ortho-to-para ratio. 
Based on ISO/SWS observations, Neufeld, Melnick \& Harwit (1998) 
estimate a value of 1.2 averaged over all the region, while Neufeld et al. (2006) 
derive values from 0.4 to 2 in knots C and E,K, which they interpret as a legacy of an earlier stage when 
the gas had reached equilibrium at a temperature $\la$ 90 K.
Disregarding the ratio involving the
5.5\,$\mu$m line, whose image seems to be affected by calibrations problems (as also suggested by the Boltzmann diagram, see 
Fig.\ref{fig:zones_diag}), we measure ortho-to-para ratios of 0.9, 1.3 and 2.3, along the jet direction, from knots C to A.  
These values, which are in good agreement with both the estimates of Neufeld, Melnick \& Harwit (1998) and Neufeld et al. (2006)
could be interpreted in terms of subsequent shock heating events due to an episodic jet.
     
\begin{figure}[h!]
\centering
\includegraphics[width=8cm,angle=0]{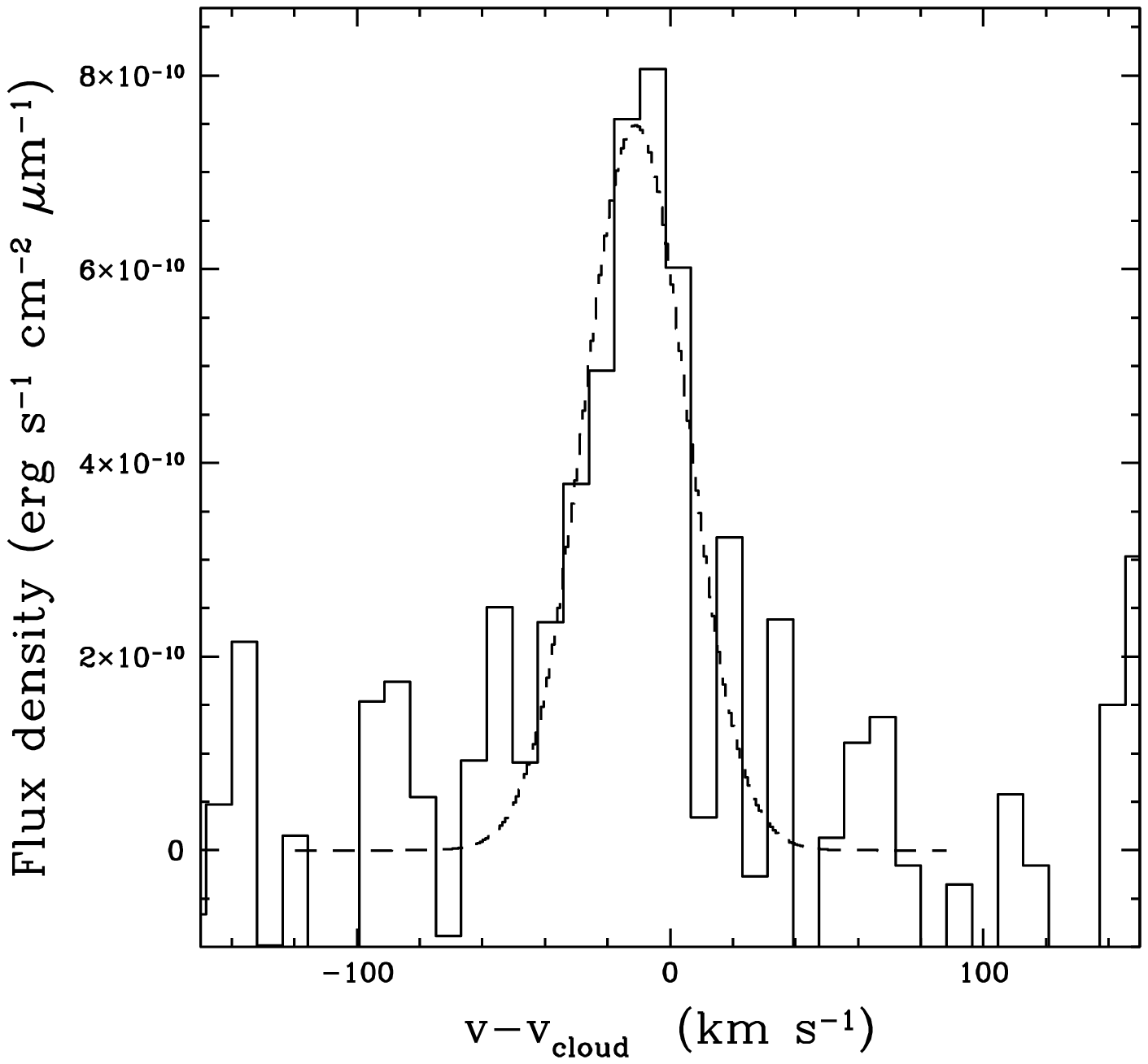}
\caption{\label{fig:FP} LWS-FP spectrum of the [{\oi}]63 $\mu$m line. The Gaussian fit is shown 
as a dotted line.}
\end{figure}

\begin{table*}
\caption{\label{tab:isocam}Photometry of (0-0) lines observed with ISOCAM.}
\begin{center}
\begin{tabular}{cccccc}
\hline\\[-5pt]
Term          &  $\lambda$& \multicolumn{3}{c}{F$\pm$$\Delta$F (10$^{-13}$erg\,cm$^{-2}$\,s$^{-1}$)}   \\
              &           &      C,L           &    B,M	         &     A   	    &     E,K,J        \\
	      &   ($\mu$m)& (135 arcsec$^2$)   & (207 arcsec$^2$)& (155 arcsec$^2$) & (173 arcsec$^2$) \\
\hline\\[-5pt]
0-0S(2)       &  12.3     & 5.5$\pm$0.2        &  5.5$\pm$0.3	 & 1.4$\pm$0.2	    & 5.9$\pm$0.2      \\	
0-0S(3)       &  9.6      & 3.9$\pm$0.3        &  6.1$\pm$0.4	 & 3.0$\pm$0.3	    & 7.0$\pm$0.4      \\   
0-0S(4)       &  8.0      & 3.4$\pm$0.3        &  4.3$\pm$0.3	 & 1.2$\pm$0.3	    & 3.6$\pm$0.4      \\   
0-0S(5)       &  6.9      & 3.6$\pm$0.3        &  6.5$\pm$0.3	 & 3.1$\pm$0.3	    & 7.7$\pm$0.4      \\   
0-0S(6)       &  6.1      & 1.9$\pm$0.3        &  3.9$\pm$0.4	 & 1.4$\pm$0.3	    & 3.5$\pm$0.4      \\	
0-0S(7)       &  5.5      & 1.3$\pm$0.4        &  2.2$\pm$0.6	 & 1.3$\pm$0.4	    & 3.3$\pm$0.5      \\   
\hline\\[-5pt] 
\end{tabular}
\end{center}
Note: Due to transient effects at low signal levels, the fluxes could be underestimated by a factor 1.3-1.6 
(relative fluxes, excitation temperatures, and o/p ratios would remain unaffected).
\end{table*}

\subsubsection{Spectroscopy}
\begin{table*}
\caption{\label{tab:lws} Far infrared lines observed in HH54.}  \small
\begin{center}
\begin{tabular}{cccc}
\hline\\[-5pt]
Term                                       & $\lambda$$_{obs}$    &  $\lambda$$_{vac}$           & $F\pm\Delta~F$          \\
                                           & ($\mu$m)             & ($\mu$m)                     & (10$^{-13}$ erg s$^{-1}$ cm$^{-2}$)\\
\hline
$[{\ion{O}{i}}]$\,$^3P_1$-$^3P_2$          &  63.18               & 63.18                        & 102$\pm$   2           \\
o-H$_2$O\, 2$_{21}$-1$_{10}$               & 108.07$^b$           & 108.07                       & $<$ 1.2                \\
CO\,23-22 + o-H$_2$O\, 4$_{14}$-3$_{03}$   & 113.54               & 113.458-113.537              & 1.3$\pm$0.4            \\
OH\,$^2\Pi_{3/2,5/2}-^2\Pi_{3/2,3/2}$      & 119.45               & 119.23-119.44                & 3.0$\pm$0.3            \\
CO\,20-19$^a$                              & 130.40               & 130.37                       & 0.9$\pm$0.3            \\
CO\,19-18                                  & 137.20$^b$           & 137.20                       & $<$1                   \\
p-H$_2$O\, 3$_{13}$-2$_{02}$               & 138.57               & 138.53                       & 1.0$\pm$0.3            \\
CO\,18-17                                  & 144.78$^c$           & 144.78                       & 2.1$\pm$0.3            \\
$[{\ion{O}{i}}]\,^3P_0-^3P_1$              & 145.52$^c$           & 145.52                       & 4.5$\pm$0.3            \\
CO\,17-16                                  & 153.19               & 153.27                       & 2.2$\pm$0.3            \\
$[{\ion{C}{ii}}]$\,$^2P_{3/2}$-$^2P_{1/2}$ & 157.72               & 157.74                       & 9.0$\pm$0.4            \\
CO\,16-15                                  & 162.73$^c$           & 162.81                       & 3.9$\pm$0.2            \\
OH\,$^2\Pi_{1/2,3/2}-^2\Pi_{1/2,1/2}$$^a$  & 163.04$^c$           & 163.12-163.40                & 1.5$\pm$0.2            \\
CO\,15-14                                  & 173.63$^b$           & 173.63                       & 6 $\pm$1               \\
o-H$_2$O\, 3$_{03}$-2$_{12}$               & 174.63$^b$           & 174.63                       & 2$\pm$1$^d$            \\
o-H$_2$O\, 2$_{12}$-1$_{01}$               & 179.53               & 179.53                       & 7$\pm$1                \\
CO\,14-13                                  & 185.85               & 186.00                       & 6$\pm$1                \\
\hline
\end{tabular}
\end{center}
Notes: $^a$ Not detected by Nisini et al.(1996);
$^b$ fixed parameter; $^c$ deblended lines; $^d$ signal to noise ratio less than 3.
\end{table*}

The 43-197 $\mu$m spectrum of HH54 was obtained with the Long Wavelength Spectrometer (LWS) aboard ISO
and reported by Nisini et al. (1996) and Liseau et al. (1996). The spectrum was taken at the beginning
of the ISO mission (orbit \#92) and originally processed with a preliminary version of the Offline Processing Software (OLP).
Recently, ISO telemetry was reprocessed (with the OLP10.1) and the resulting Relative Spectral Response Function
(RSRF) was higher than in the earlier versions; here we report the analysis of this upgraded version 
of the spectrum.
The observation was carried out with the LWS AOT01 ($\lambda/\Delta\lambda$ $\sim$ 200), adopting the ``fast scanning''
option and an oversampling of 4 times the spectral resolution element. The flux calibration, based on Uranus observations,
has an uncertainty of up 30\%, while the accuracy on the wavelength calibration is about 25\% of the spectral element
(0.29 $\mu$m and 0.60 $\mu$m for the short [43-90 $\mu$m] and long [90-197 $\mu$m] wavelength range, respectively).
Post-pipeline processing, which consists of glitch removal, averaging of different spectral scans, and correction
for interference fringes, was performed with the ISO Spectral Analysis Package
\footnote{The ISO Spectral Analysis Package ISAP is a joint development by the LWS and SWS Instrument Teams and Data Centers.
Contributing institutes are CESR,IAS,IPAC,MPE,RAL and SROAN.} (ISAP), v.2.1. 
In comparison with the values reported by Nisini et al. (1996), the new line fluxes, given
in Table \ref{tab:lws}, are systematically 30-40\% lower.  In addition, two more lines (CO\,20-19 at 130.4 $\mu$m
and OH at 163 $\mu$m)  have been detected, but we do not observe the CO\,19-18 line
at 137 $\mu$m, which was just detectable in the old version of the spectrum (shown by Liseau et al. 1996).
The low resolution data have been complemented by the intermediate resolution spectrum of the [{\oi}] 63 $\mu$m line
(see Fig. \ref{fig:FP}). This was accomplished by coupling the LWS with a Fabry-Perot interferometer, thus achieving
a resolution of $\approx$8500 (Swinyard et al., 1996). The line, detected with a signal-to-noise ratio of about five,
peaks at v - v$_{cloud}$ = -12 km s$^{-1}$, similar to that of the 2.12\,$\mu$m line.
Although this could suggest a similar origin for oxygen and molecular hydrogen emission, the 63\,$\mu$m profile is 
$\approx$ 39 km\,s$^{-1}$ wide, and the corresponding FWZI, computed on the line fit by applying the procedure of Hartigan, 
Raymond \& Hartmann (1987), is $\approx$ 80 km\,s$^{-1}$. This value can only be considered a rough estimate,
given the poor signal-to-noise ratio of the detection, nevertheless it indicates that the 63\,$\mu$m emission 
may arise in a shock with velocity larger than (or close to) the H$_2$ dissociation limit.

\section{Analysis and Discussion}

\subsection{Boltzmann diagrams}

\begin{figure*}
\centering
\includegraphics[width=15cm]{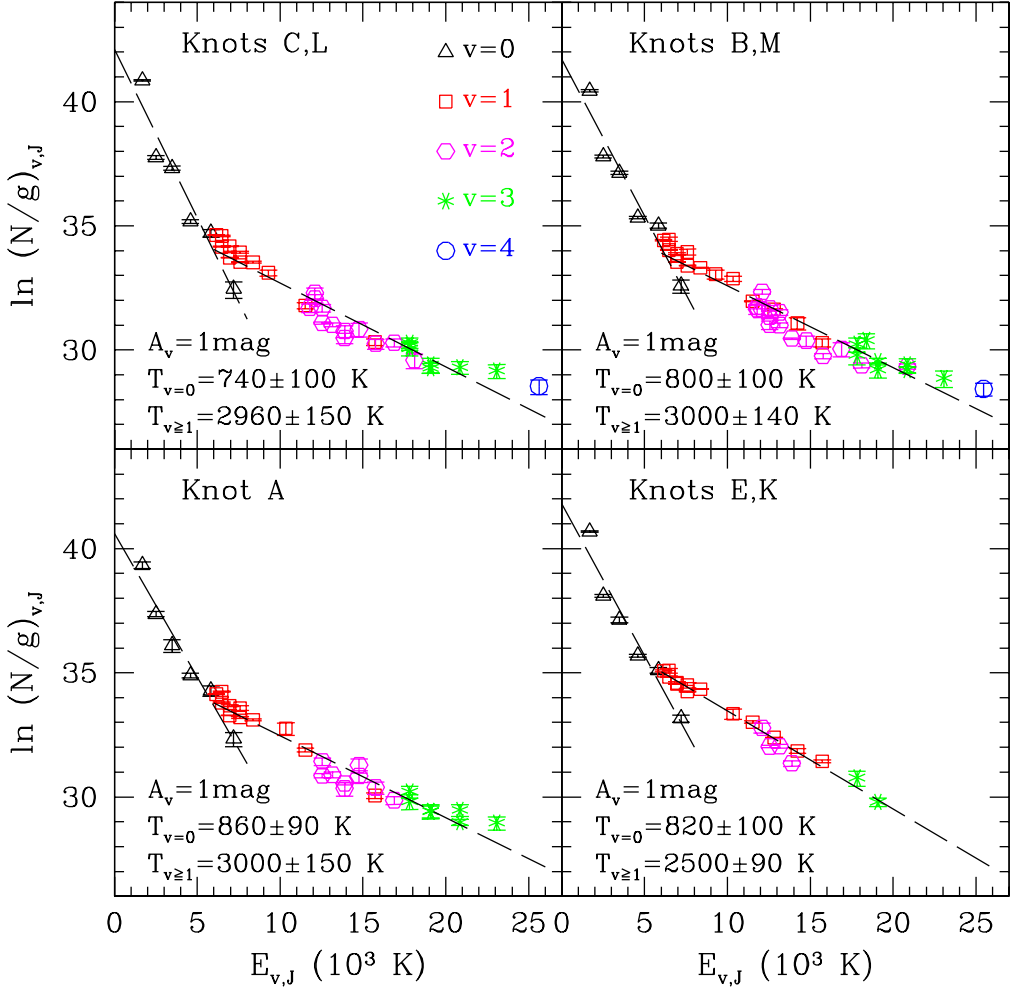}
\caption{\label{fig:zones_diag}Boltzmann diagram of the four zones identified in the (0-0) images (see text). In the upper-left panel,
symbols adopted to mark different vibrational levels are shown. In each diagram, the derived extinction, along with the
temperatures of the vibrational manifolds, are indicated.}
\end{figure*}

Analysis of the H$_2$ line emission gives rise to estimates for the reddening towards the encompassed
knots and the temperature of the emitting gas. These two quantities can be simultaneously
estimated in the framework of the Boltzmann (or excitation) diagram, 
which is a plot of ${\rm ln}(N_{v J}/g_J)$ against $E_{v J}/k$, where $N_{v J}$ (cm$^{-2}$) 
is the column density of level ($v, J$), $E_{v J}/k$ (K) its excitation energy and $g_J = (2J+1)(2I+1)$ its statistical 
weight. The nuclear spin quantum number is $I = 1$ for ortho--H$_2$ and $I = 0$ for para--H$_2$.
If the gas is thermalized at a single temperature, the points in the diagram fall onto
a straight line: a best fit (by minimization of the deviation of the data-points from the line) to the gradient provides 
the reciprocal of the gas temperature.  At the same time, under the assumption of optically thin emission, the extinction 
can be evaluated by noting that transitions from the same upper level should lie at the same 
point in the diagram (e.g. G94). Thus, having adopted the reddening law by Rieke \& Lebofsky (1985), 
we have tuned the A$_V$ value to maximize the overlap of data points corresponding to lines originating in the same
upper level. We derive A$_V$=1\,mag over all of the region.

In a shock wave there is a range of kinetic temperatures and the Boltzmann diagram forms a curve.  In order, therefore, 
to probe our observations at different temperatures, columns of the pure rotational lines are plotted in a
Boltzmann diagram together with those of near-infrared ro-vibrational lines. Since
the region emitting in 0-0 lines has been separated in four distinct zones (as described in Section 2.2.1), 
we have constructed four Boltzmann diagrams, plotted in Fig. \ref{fig:zones_diag}; in each zone, 
the column densities of the ro-vibrational lines have been computed by assuming an emitting region defined from 
the 3$\sigma$ contours of the 2.12\,$\mu$m line. 
The similarity of the four Boltzmann diagrams in Fig. \ref{fig:zones_diag} implies that the excitation conditions are broadly 
the same for all of the knots in the HH54 region, at least at the spatial resolution ($\approx$ ten arcsec) characteristic of the 
four zones. Each diagram indicates the presence of two temperature components. 
A 'cold' component at T$\approx$ 800 K is traced by the 0-0 lines: this is in substantial agreement  with that measured by Neufeld,
Melnick \& Harwit (1998) and Cabrit et al.(1999) on the basis of SWS and CAM observations ($\approx$ 650 K). 
All the other lines appear thermalized at about $\sim$ 3000 K from north to south (knots C to A) and at $\sim$ 2500 K on the wing;
noticeably, this latter result could simply reflect the lack along the wing direction of spectroscopic data in the J band, 
where H$_2$ lines with high excitation energy (i.e. those tracing hotter temperature components), are located (Giannini et al., 2002).
In the plot, the ortho-to-para ratio was assumed equal to the statistical equilibrium value of three: any significant
departure from this value should result in a systematic displacement
of the ortho with respect to the para levels. While such behaviour is observed among the $v$=0 lines (see also Table \ref{tab:isocam}), 
no significant deviations from the equilibrium value are appreciable in the data points tracing
the hotter gas components (those with  $v$$\ge$1).This can more easily be seen by taking ratios of fluxes of bright ortho and para lines belonging to the same 
manifold (e.g. those among 1-0 and 2-1 lines, see Table \ref{tab:fluxes}), for which the extinction correction is 20\% at worst. 
  
\subsection{Models of Molecular Emission}\label{h2}

In the previous Section we showed that the observed emission is quantitatively similar in both the jet and the wing of HH54, indicating common 
excitation mechanisms and similar physical conditions for the whole region.
Accordingly, we can examine together the H$_2$ emission with the emission of the other molecular species (CO, H$_2$O and OH),
which are observed with a beam encompassing the whole HH54 region. Furthermore, we have taken the zone encompassing knots B and M 
(hereafter the ``B-zone''), in which \h \, is emitted over the widest range of energies observed, to be representative of the excitation conditions in HH54.
To explore the diagnostic capabilities of the observed lines in disentangling different shock structures and physical parameters, 
we have analyzed the observations in the context of a full range of shock models, including steady-state non-dissociative (C-type) and 
dissociative (J-type) shocks and quasi-steady J-type shocks with magnetic precursor 
(denoted C, J and C+J, respectively, for brevity). For each type of shock we have proceeded in the following way: {\it i)} the \h \, column 
densities in the B-zone were fitted using a shock model, which is described in Section \ref{sec:H2}; {\it ii)} temperature and density profiles
along with local velocity gradient predicted by each best-fitting model were used as input parameters for a Large Velocity Gradient (LVG) 
code to reproduce the H$_2$O, CO and OH FIR line emission (see Section \ref{sec:LVG}). At this step we can both test the 
overall consistency of the selected models and also identify the FIR lines most useful in 
diagnosing the shock physics; {\it iii)} finally, in order to determine if atomic/ionic and molecular emission are excited under the 
same conditions, the {\oi},\,{\ci},\,{\s} and {\fe} lines fluxes predicted by the best shock model found for the molecular lines were 
compared with the NIR observations (Section \ref{sec:ion}).

\subsubsection{H$_2$ Emission}\label{sec:H2}

\begin{figure}
\centering
\includegraphics[width=8cm]{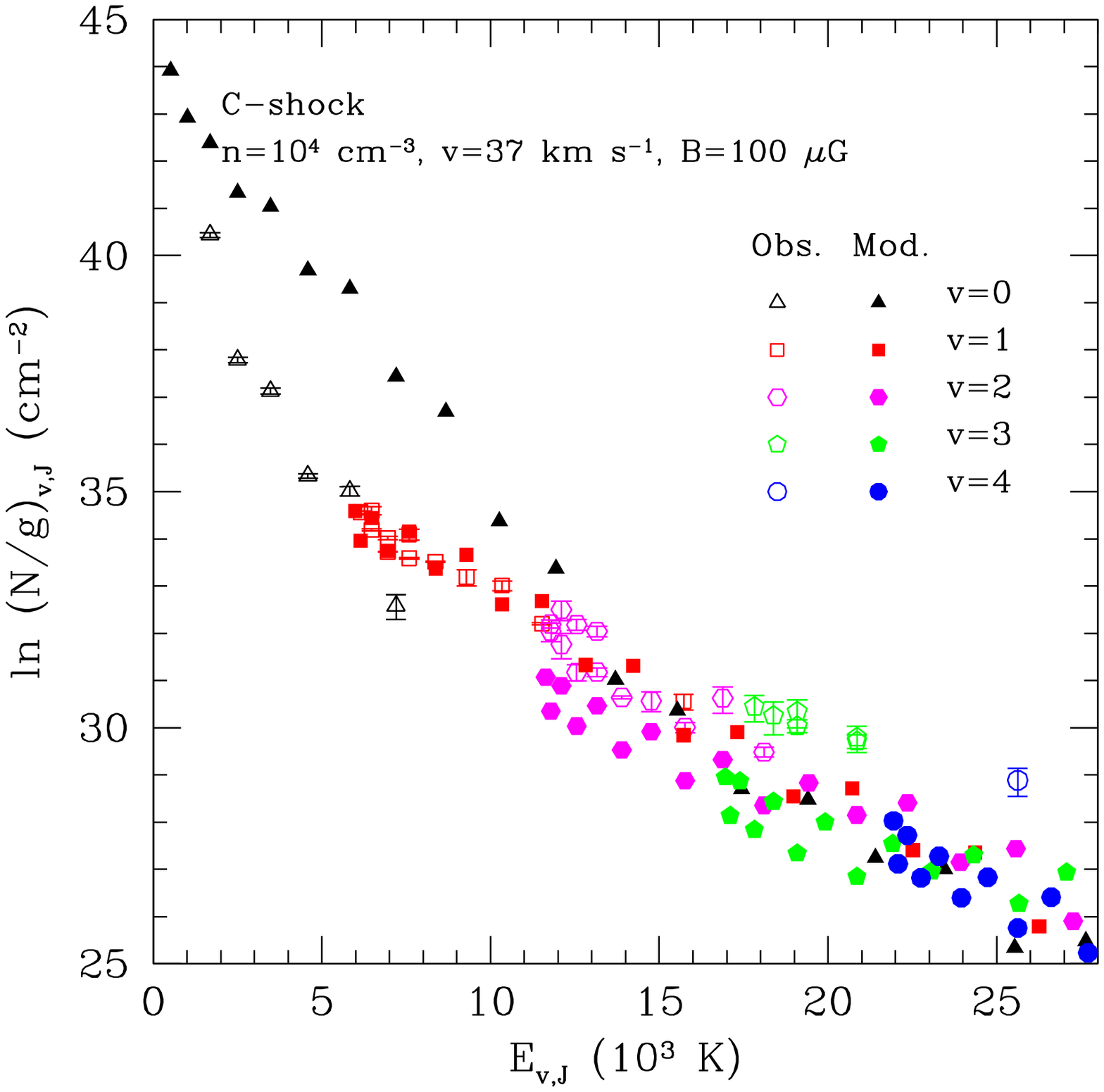}
\includegraphics[width=8cm]{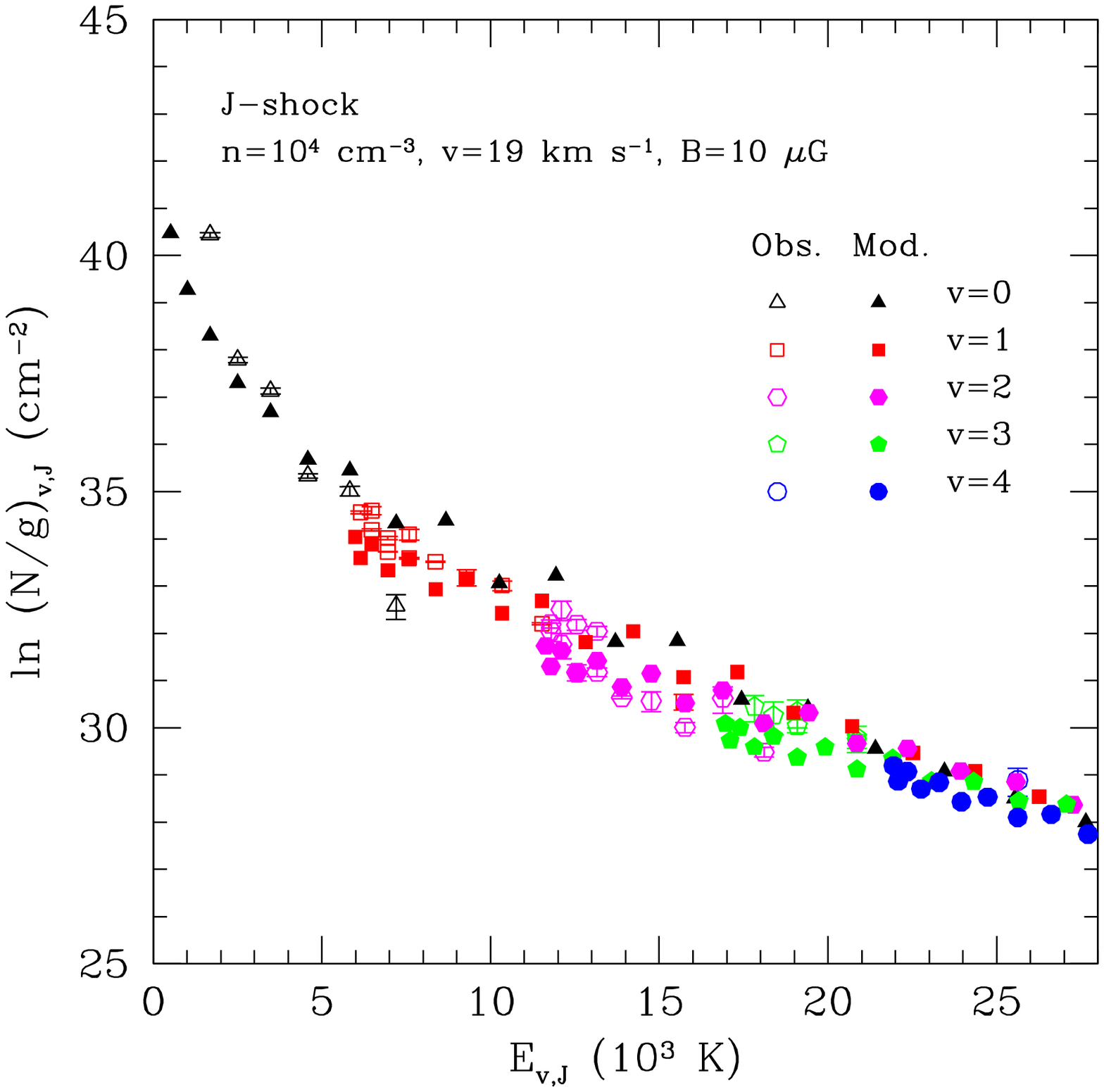}
\includegraphics[width=8cm]{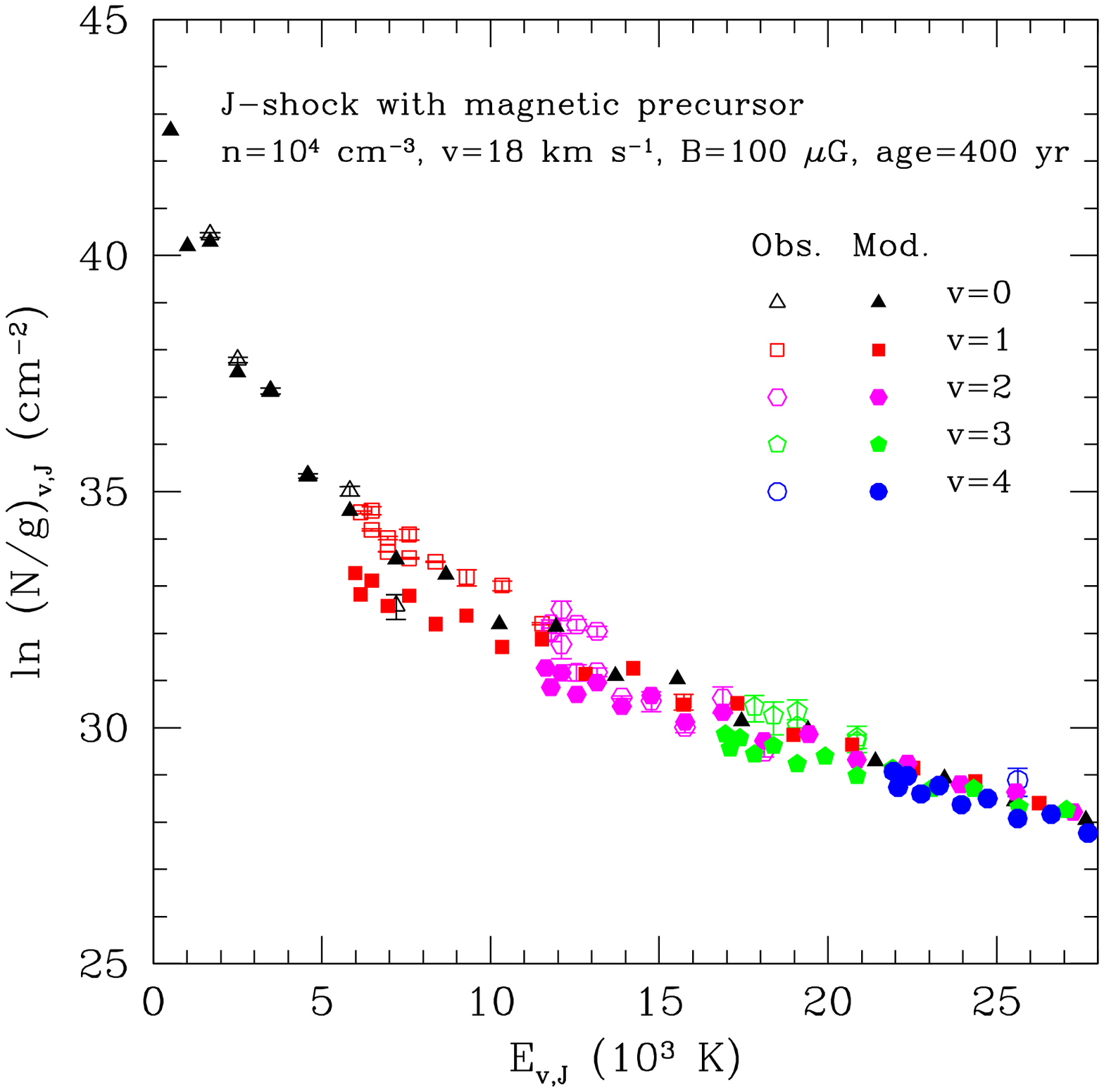}
\caption{\label{fig:H2fit} H$_2$ Boltzmann diagrams of the B-zone compared with the predictions for a steady-state
C-type and J-type shocks (top and middle panels) and for a J-type shock with a magnetic precursor (bottom panel). The
shock model parameters are given in each panel and each vibrational state is labelled with a symbol, as indicated.}
\end{figure}

\begin{figure}
\centering
\includegraphics[width=7.7cm]{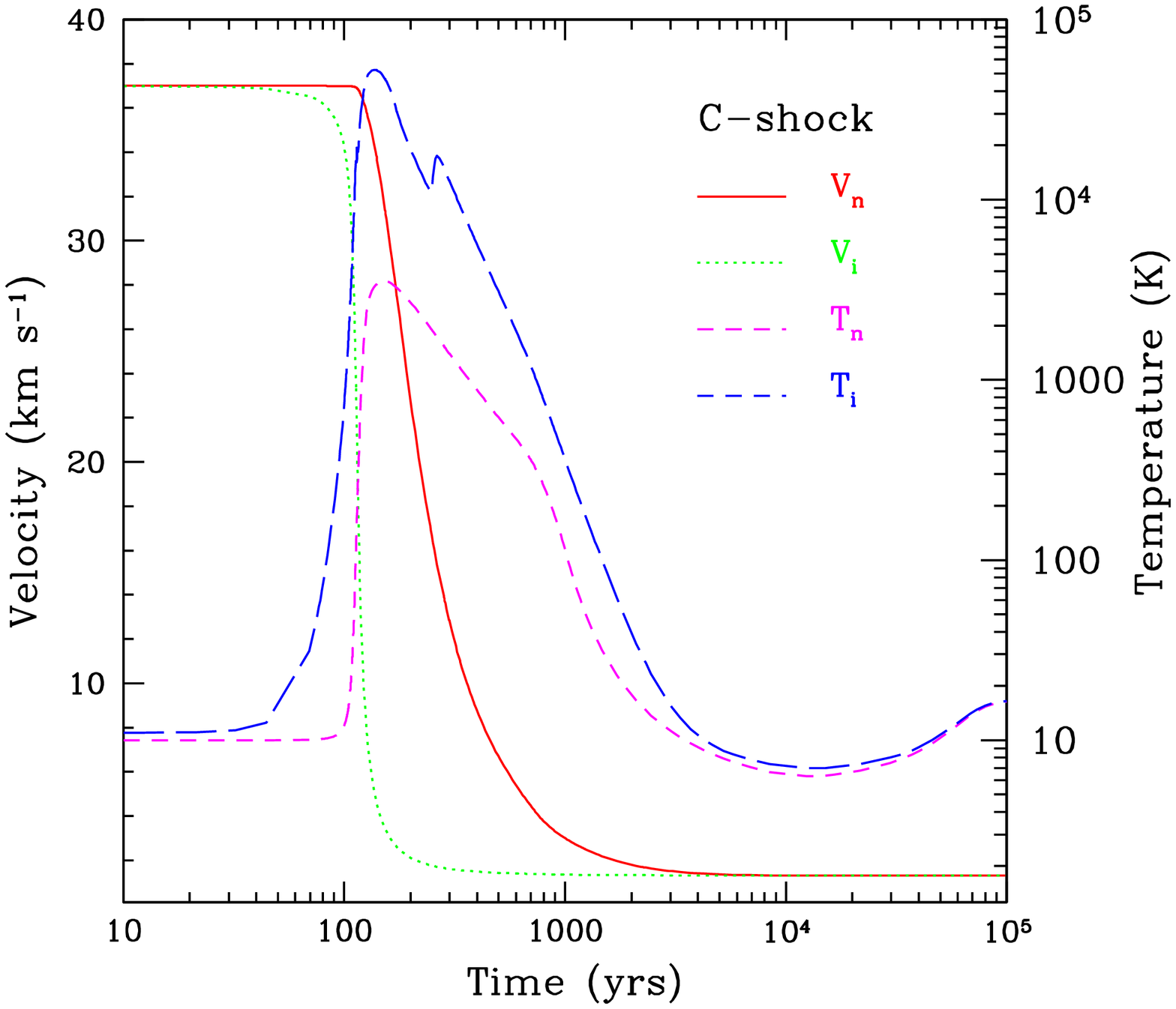}
\includegraphics[width=7.7cm]{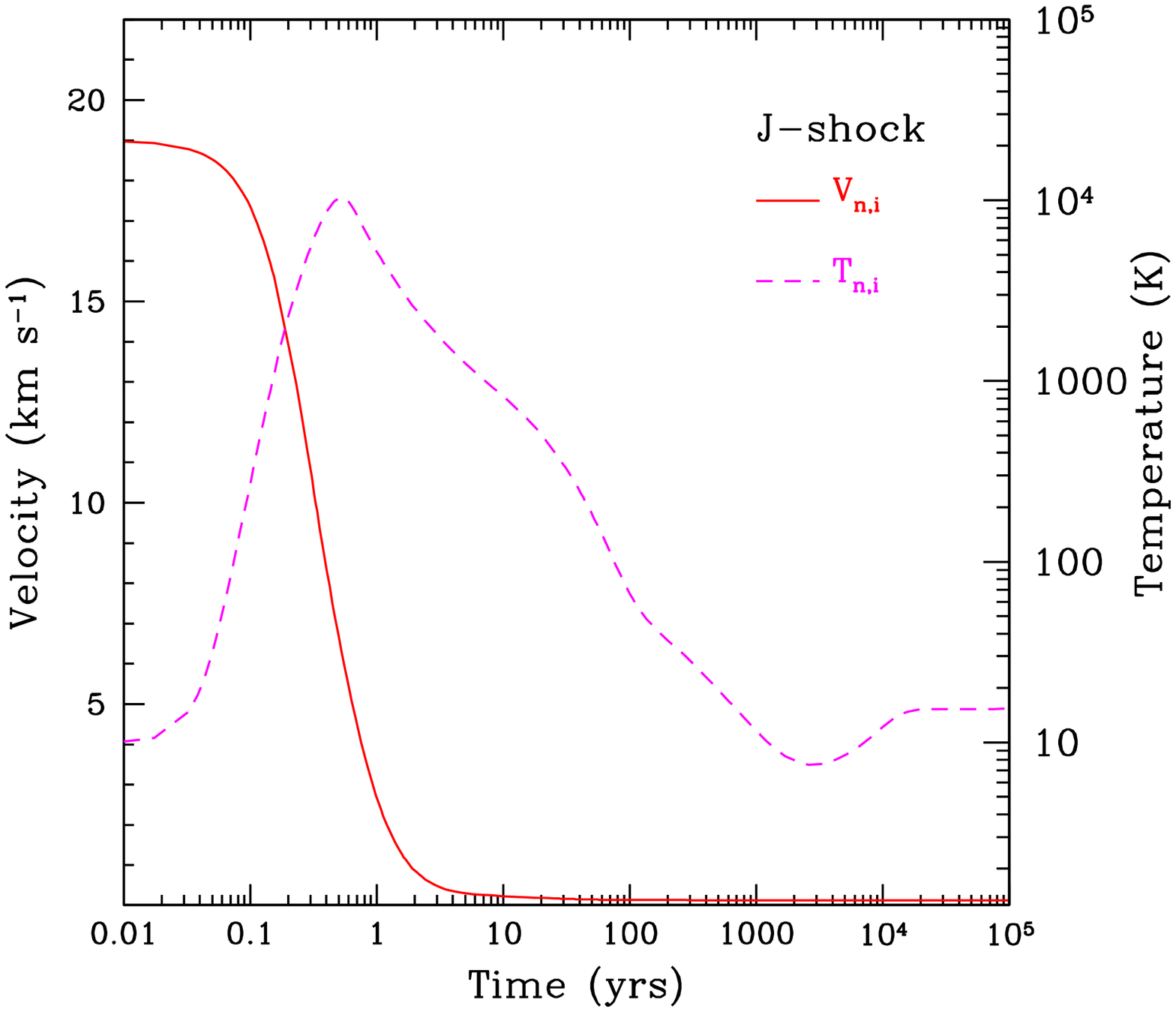}
\includegraphics[width=7.7cm]{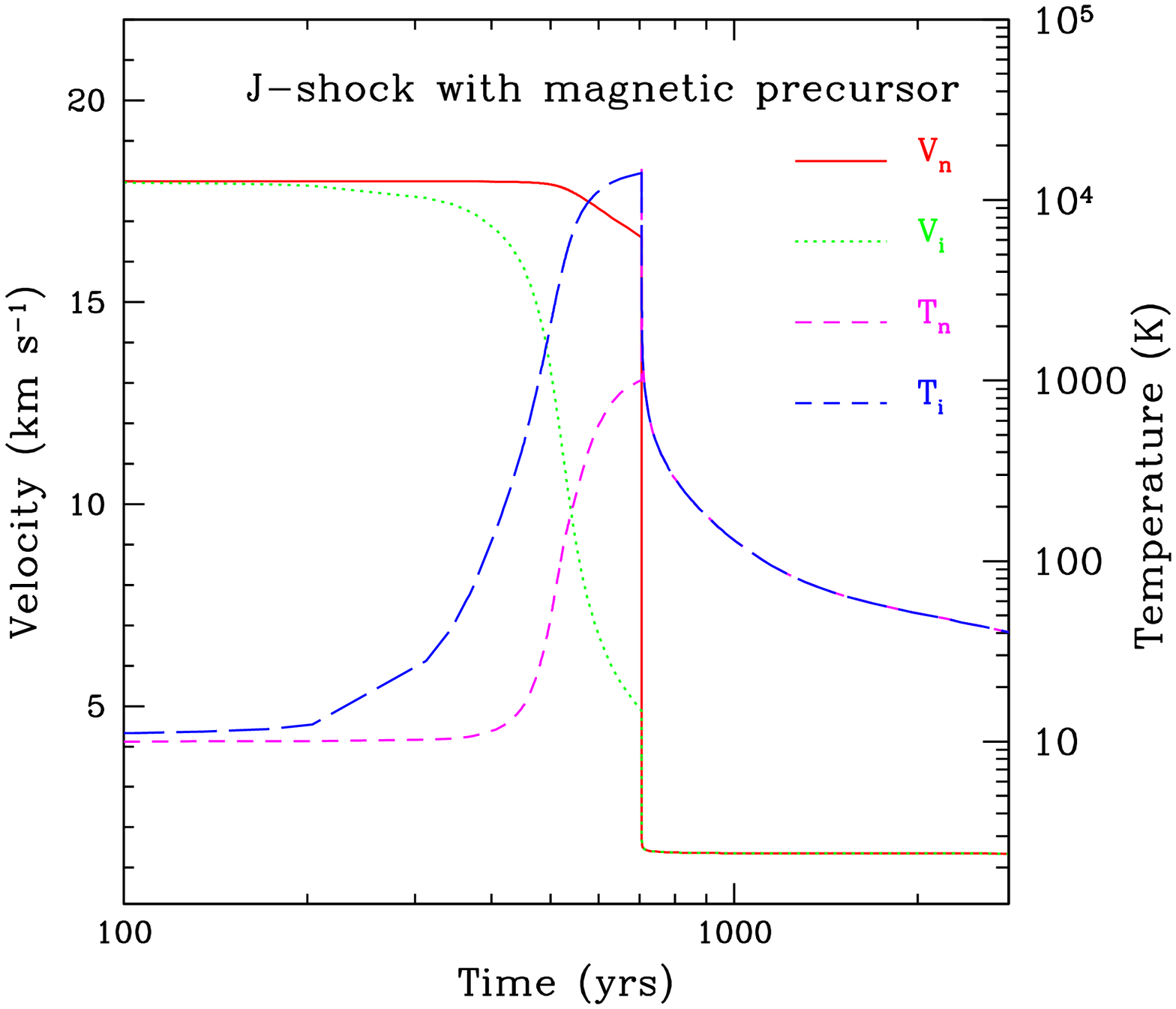}
\caption{\label{fig:structure} Velocity and temperature profiles from the models of the best-fitting steady-state C-type and J-type shock (top and
middle panels) and the best-fitting J-type shock with a magnetic precursor (bottom panel). The velocity is expressed in the shock frame 
and the pre-shock gas is to the left.}
\end{figure}

\begin{figure}
\centering
\includegraphics[width=7.7cm]{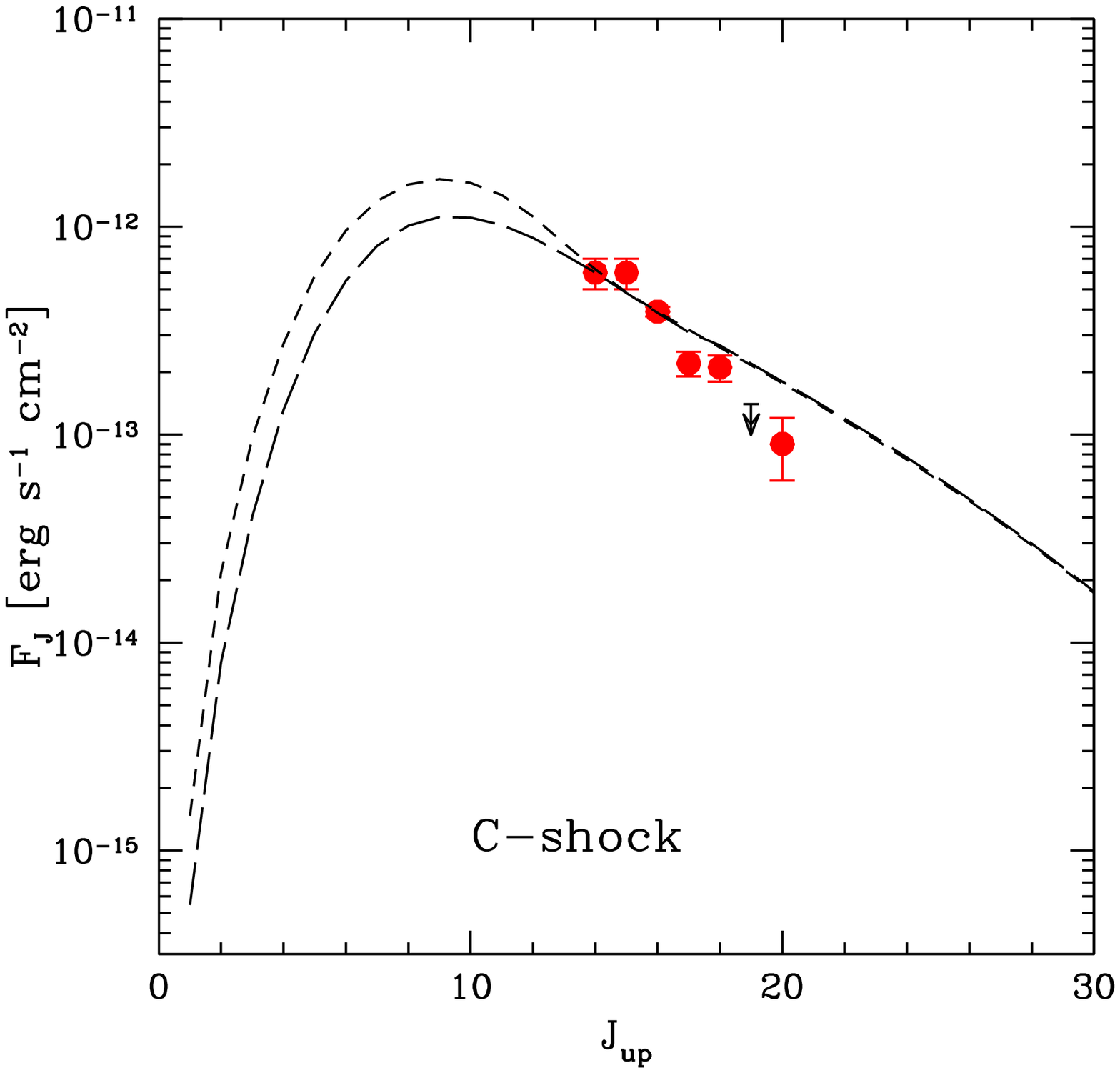}
\includegraphics[width=7.7cm]{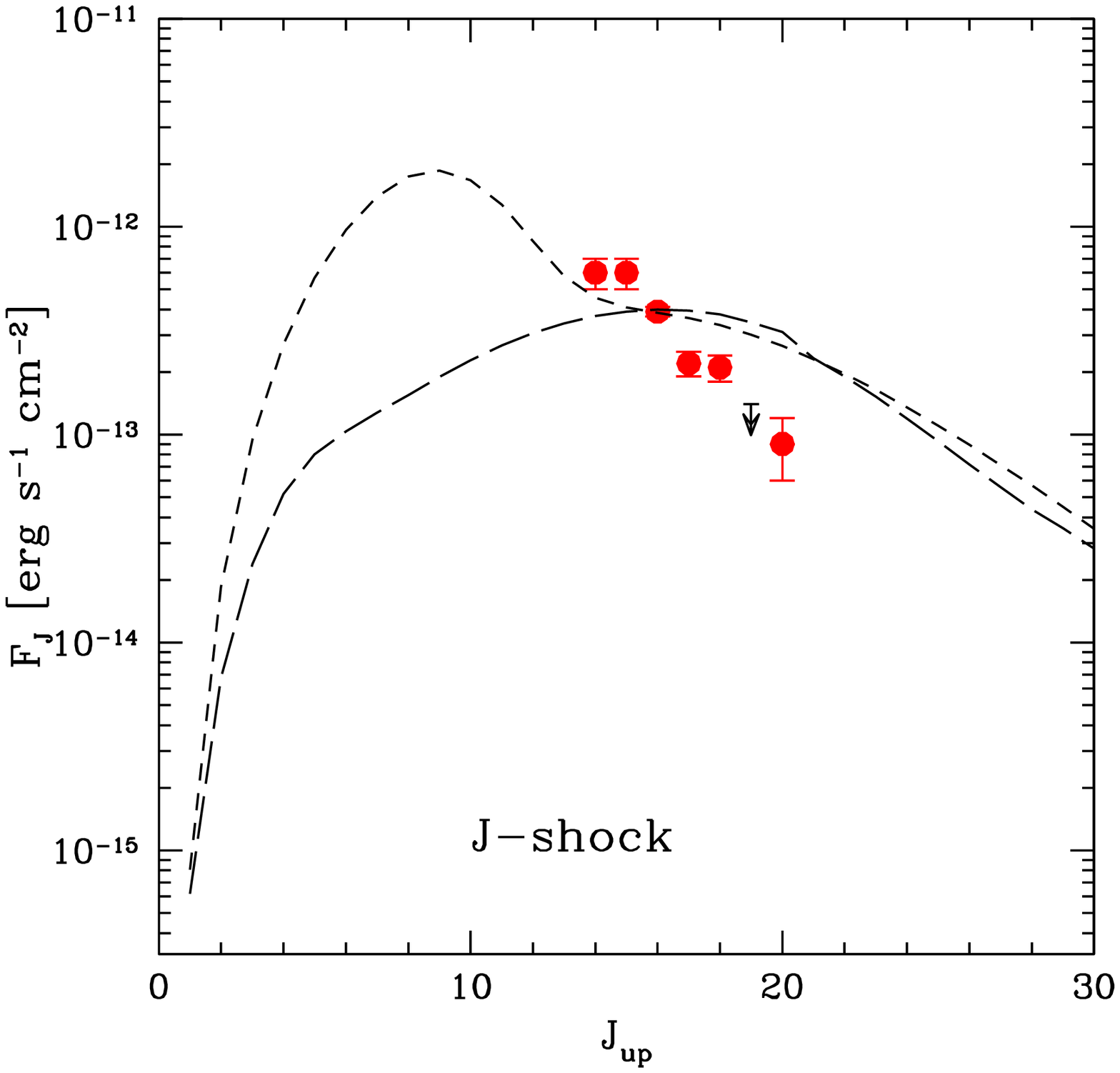}
\includegraphics[width=7.7cm]{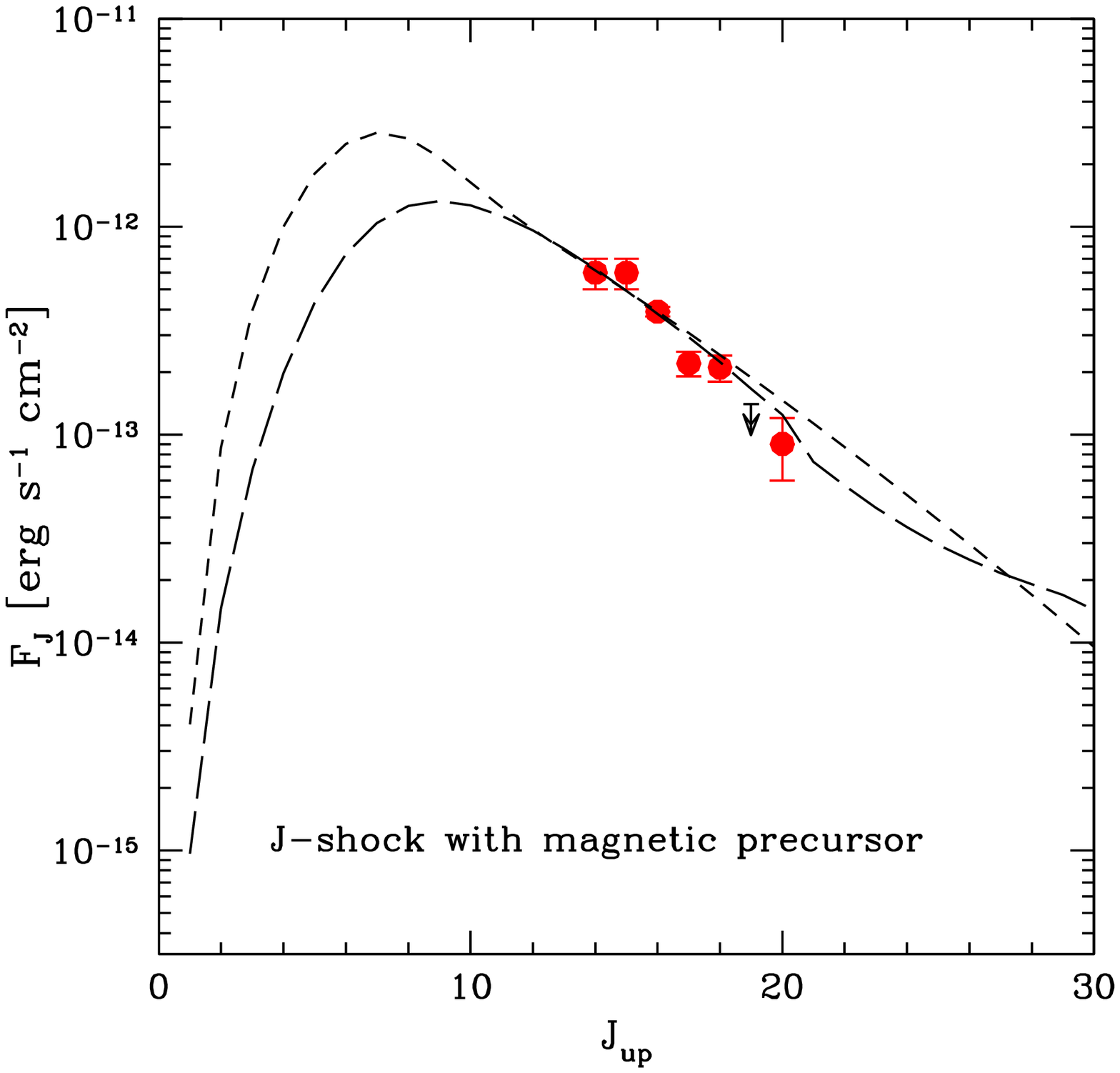}
\caption{\label{fig:CO} {\it Top:} CO line fluxes as a function of the rotational quantum number J$_{up}$.
Continuous lines give the fluxes predicted by the C-shock model by assuming the collisional parameters
of McKee (1982) (short-dashed) and of Flower (2001) plus Sch\"oier et al. (2005) (long-dashed). 
The error bars represent the 1$\sigma$ uncertainty. The downward arrow indicates a 3$\sigma$ upper limit; 
{\it middle:} as for the top panel but for the 
steady-state J-shock; {\it bottom:} as for the top panel but for the J-shock with a magnetic precursor.}
\end{figure}

\begin{figure}
\centering
\includegraphics[width=9cm]{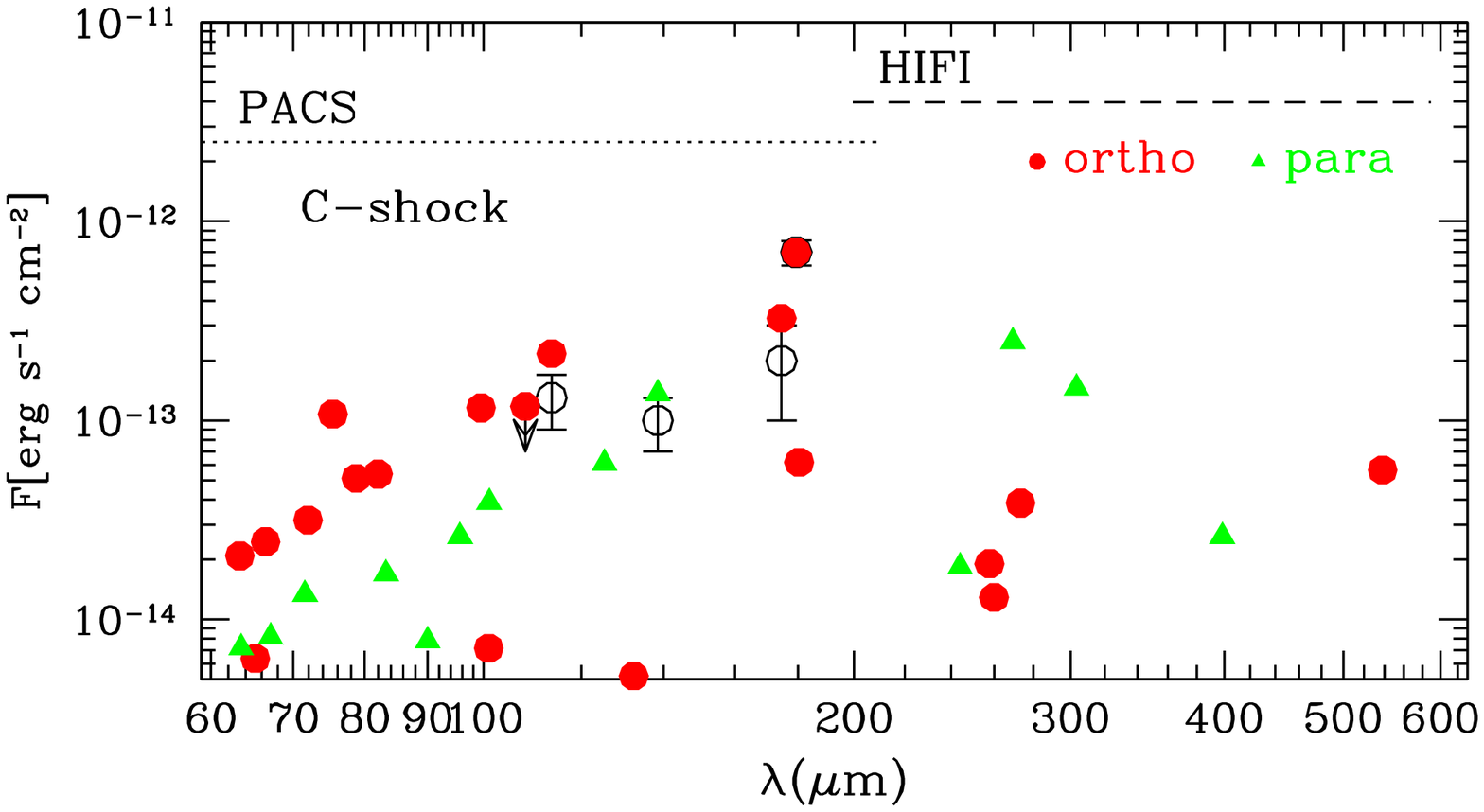}
\includegraphics[width=9cm]{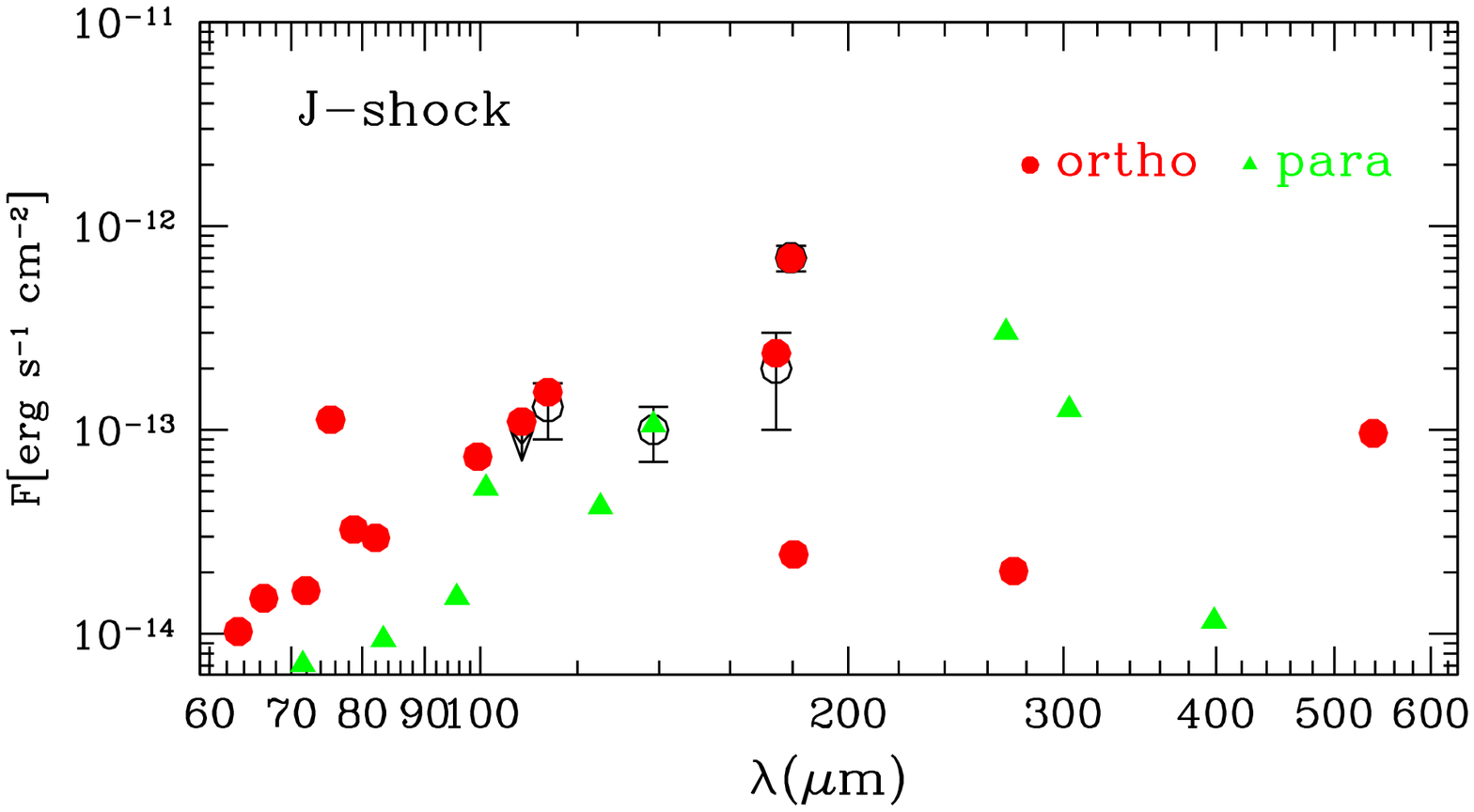}
\includegraphics[width=9cm]{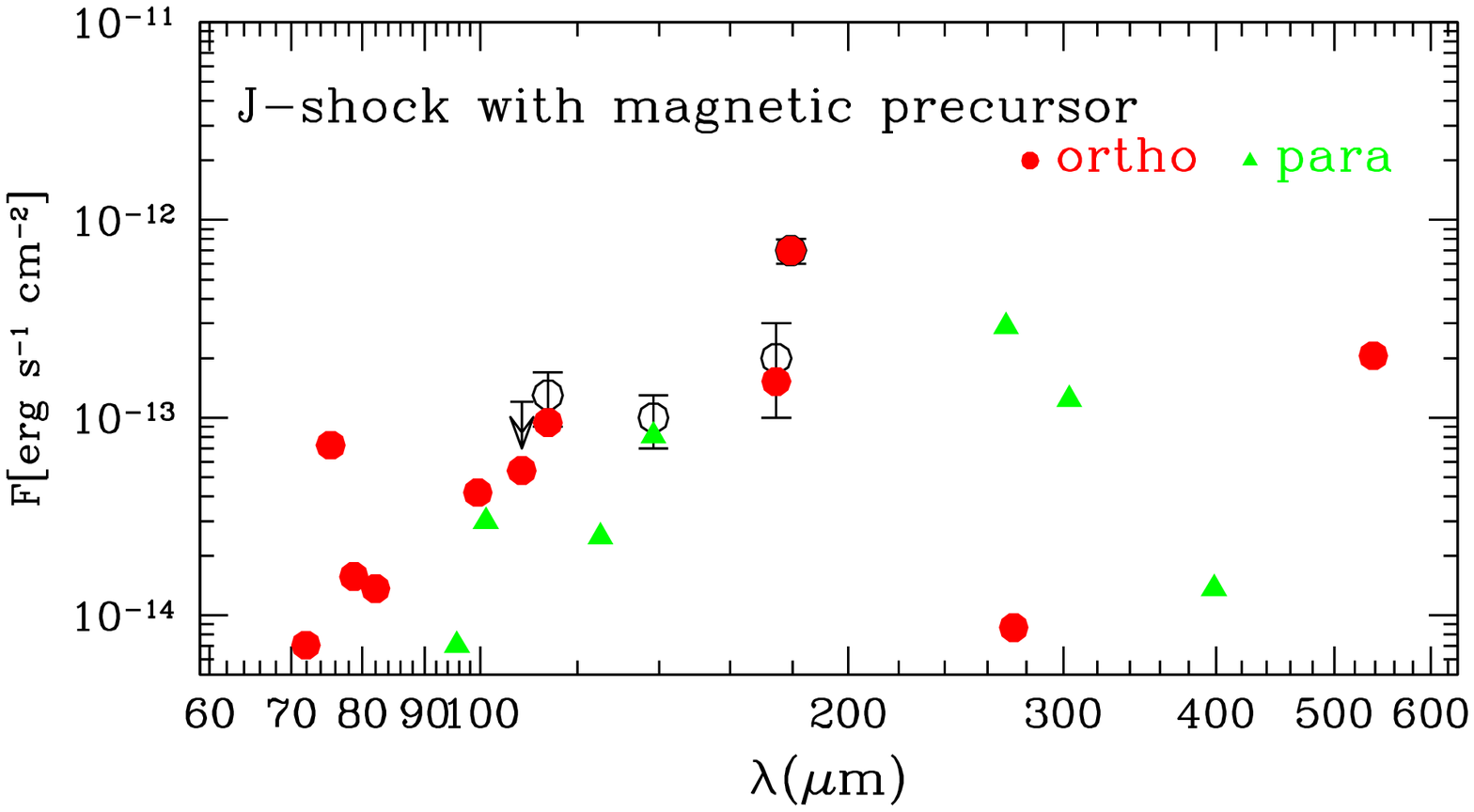}
\caption{\label{fig:h2O} {\it Top panel:} H$_2$O line fluxes as a function of wavelength.
The LWS spectral range extends up to 200 $\mu$m, but we report in the figures all the lines
falling in the range open to Herschel observations.
Data points with the relative 1$\sigma$ uncertainties are represented with open circles, while the ortho- and
para-H$_2$O fluxes predicted by the C-shock model of Tab. \ref{tab:param} are shown with filled circles
and triangles, respectively. The downward arrow indicates a 3$\sigma$ upper limit. The wavelength range covered by PACS and HIFI is indicated,
as well. {\it Middle panel:} as in the top panel 
but for the steady-state J-shock model. {\it Bottom panel:} as in the top panel but for the J-shock 
with a magnetic precursor model.}
\end{figure}

The shock code, \textsc{MHD$\_$VODE}, has been described by Le Bourlot et al. (2002) and Flower et al. (2003), while the method used 
to calculate the H$_2$ emission has been explained by Giannini et al. (2004) and M$^\mathrm{c}$Coey et al. (2004).  Here, only the 
aspects most relevant to this study are reviewed.  

The code solves for one--dimensional, planar, multi--fluid flow and can simulate both steady-state J- and C- type shock waves and
quasi-steady J-type shocks with magnetic precursor. The differential equations that
determine the abundances of the chemical species are solved in parallel
with the magnetohydrodynamical conservation equations for the neutral,
positively and negatively charged fluids. 

The calculations 
for the distribution of population among 
150 H$_2$ ro-vibrational levels (i.e. up to an energy of 3.9$\times 10^4$ K) take into account collisional 
excitation and de-excitation, spontaneous radiative decay, collisional dissociation and ionization, and reformation of 
H$_2$ on grains, and are solved in parallel with the chemical and dynamical conservation equations. This approach is essential 
to ensure the accuracy of the computed H$_2$ column densities, because the level populations do not respond instantaneously to changes in 
the physical state of the gas.   

The column density of a level is sensitive to the pre-shock density, the shock velocity, the age of the shock and the 
strength of the magnetic field. The response of the level populations to these variables, which are parameters of the code, 
can be seen in the excitation diagram.
If the gas is in local thermodynamic equilibrium (LTE), the Boltzmann plot is a smooth curve, with little
scatter of the points about the median line. Departures from LTE enhance the scatter.  Increases in the shock velocity, $v_\mathrm{s}$, 
and the pre-shock density, $n_\mathrm{H}$, act to increase the rate of 
collisions that vibrationally excite the H$_2$ molecules and thus decrease the departure from LTE. The presence of a 
magnetic field acts to dampen and broaden the shock wave; therefore, an increased 
magnetic field strength enhances the departure from LTE.  The magnetic field is assumed to scale by a factor $b$ with 
the square root of the gas density and is expressed as $B$($\mu$G)$ = b[n_\mathrm{H}$(cm$^{-3}$)]$^{0.5}$.
The assumed dependence of $B$ on $n_{\rm H}$ derives from considerations of energy equipartition: the magnetic
energy density is $\propto B^2$, whereas the thermal energy density is $\propto n$. We note that the shock front is assumed
to be planar, and the parameter $b$ scales the component of the magnetic field parallel to the shock front (and
perpendicular to the flow direction), consonant with the assumed dependence of $B$ on $n_{\rm H}$. Thus, 
$n_{\rm H} = 10^4 \,cm^{-3}$, $b=1$ corresponds to a transverse magnetic field strength $B = 100 $~$\mu $G.\\
The evolutionary age of the shock wave has a marked effect on the excitation diagram. The younger the object, 
the closer is the shock wave to being pure J-type and the higher are the column densities of the more highly excited 
ro-vibrational levels.  Furthermore, the populations of these levels tend to be closer to LTE.  In a J-type shock with 
a magnetic precursor, the levels of $v = 0, 1$ are populated principally in the precursor and exhibit greater 
departures from LTE than the higher levels. Furthermore, most of the population remains in the $v = 0$ ground vibrational 
state due to radiative cascade from excited vibrational states.

The code was systematically applied over a grid of the above parameters for each category of shock (C, J and C+J) as follows: 
10$^3$ cm$^{-3}$ $\leq n_{\mathrm{H}} \leq$ 10$^5$ cm$^{-3}$; 
1 km s$^{-1}$ $\leq v_{\mathrm{s}} \leq v_{\mathrm{crit}}$ (in the case of C-type shocks), where $v_{\mathrm{crit}}$ is 
the shock velocity at which a steady-state C-type shock becomes discontinuous due to H$_2$ dissociation (Le Bourlot et al., 2002); 0 $\mu$G 
$\leq B \leq$ 500 $\mu$G; and, for ages greater than that at which a J-type shock with a magnetic precursor becomes 
indistinguishable from a J-type shock wave (approximately 100 yrs for $n_{\mathrm{H}} = 10^4$ cm$^{-3}$ and 
$v_{\mathrm{s}}=$ 18 km s$^{-1}$).  

From rotational \h \, transitions observed towards HH54, Neufeld et al. (1998) derived a pre-shock ortho-to-para ratio of $\le$0.2. 
We investigated the effect of the initial ortho-to-para ratio on the Boltzmann diagram and found that, indeed, a value of $\le$0.2 is 
required in order to reproduce the populations in the v=0 levels. If the maximum neutral temperature is sufficiently high (exceeding
1300 K for C-type shocks. or 10$^4$ K forJ-type shocks) then the local ortho-to-para ratio attains the statistical equilibrium value
of 3 (Wilgenbus et al., 2000). This condition is met in the models considered (see Fig.7) and so the v$>$0 levels, which are populated
mostly in the hot gas behind the shock front, are insensitive to the initial ortho-to-para ratio.

\begin{table}
\caption{\label{tab:param}Parameters of shock models selected, for each category of shocks, as best fits to the H$_2$ emission.}
\begin{center}
\begin{tabular}{ccccc}
\hline\\[-5pt]
Shock type & $v_{\mathrm{shock}}$  & $n_{\mathrm{H}}^a$  & Age  & $B^b$ \\
           & (km s$^{-1}$)         & (cm$^{-3}$)         & (yr) & ($\mu$G) \\
\hline\\[-5pt]
C   &37 & 10$^4$ & --- & 100 \\
J   &19 & 10$^4$ & --- & 10  \\
C+J &18 & 10$^4$ &  400 & 100\\  
\hline\\[-5pt]
\end{tabular}
\end{center}
$^a$ $n_{\mathrm H}$ = $n$(H) + 2$n$(H$_{2}$)\\
$^b$ $B$($\mu$G)$ = b[n_\mathrm{H}$(cm$^{-3}$)]$^{0.5}$, $b=1$ for the C and C+J shock models and 0.1 for the J shock model.\\
\end{table}

For the steady C and J shock models, we have integrated line fluxes out to a flow time of 10$^5$ yr, where the post-shock 
gas has reached thermo-chemical equilibrium. For the young C+J shock, the total flow time through the shock wave should be less than 
the time to reach steady state for the C precursor, about 10$^4$ yrs (otherwise one would have a pure C-shock, not a C+J). However,
the exact flow time through the J-cooling zone following the C-precursor is not precisely constrained, as it depends on the
detailed shock history (e.g. Lesaffre et al., 2004a,b), and on any time-variability in the inflow conditions. Here we have
chosen, for illustrative purposes, to truncate our flux integration at t$_{i}$ = 3000 yrs, by this time the contribution  
to the H$_2$ and ISO FIR lines in negligible. We note that that fully time-dependent (and 2D) 
calculations are required for more detailed predictions.

The first result is that C-shock models are unable to 
reproduce the H$_2$ emission: a J-type component is required in order to produce the observed degree of thermalisation 
among the levels.  As a demonstration, the excitation diagram from the best-fitting steady-state C-type shock model 
(v$_\mathrm{s}$ = 37 km s$^{-1}$, $n_{\mathrm{H}}= 10^4$ cm$^{-3}$) is plotted along with the observations in 
Figure \ref{fig:H2fit}, top panel.  This model fits the v$=$1 levels reasonably well but overestimates the lower energy levels while 
underestimating the higher energy levels. 

On the other hand, two models were found to produce a good fit to the H$_2$ emission: a J-type shock with a magnetic
precursor of age 400 yrs
and v$_{shock}$=18 km s$^{-1}$ and a steady-state J-type shock with v$_{shock}$ = 19 km s$^{-1}$. A pre-shock 
density of 10$^4$ cm$^{-3}$ is predicted by both models.
Table \ref{tab:param} and Fig. \ref{fig:structure} report the full set of parameters and the velocity and
temperature profiles, respectively, while the 
excitation diagrams produced from these models can be compared with those derived from observations in Fig. \ref{fig:H2fit}.
The v$>$1 energy levels are fitted well by both models, because the J-component, which 
is largely responsible for the population of the higher energy levels in the case of the J-type shock with magnetic 
precursor, is essentially the same in each case (see Fig. \ref{fig:H2fit} middle and bottom panels). The C+J shock  
model reproduces the population in the v=0 levels but underestimates the v=1 levels.  
The higher degree of thermalisation arising in the steady-state J-type shock results in a greater population in the v=1 levels but this 
is at the cost of population in the v=0 levels and is still insufficient to reproduce the observations, a result which
agrees with the findings by Wilgenbus et al.(2000).

Therefore, at this step of the analysis, it can be concluded that a dissociative component is required 
to fit the H$_2$ emission, even if from the \h \, data alone, it is not 
possible to distinguish between a steady-state J-type shock and a quasi-steady J-type shock with magnetic precursor. 
It is interesting to note, however, that a model of the latter type could be used to describe the larger scale geometry
of the HH54 region. In particular, the smoothness of the bow-shape traced by the 
rotational lines (Fig.\ref{fig:isocam}), which are excited predominantly in the continuous precursor, in comparison
with the clumpiness of the ro-vibrational lines, which arise in the discontinuity and are more
affected by the time variability of the driving jet.

\subsubsection{Implications of the derived shock parameters}\label{env}
We want now briefly to analyze the environmental impact caused by the physical quantities which derive from the shock wave model. Indeed, numerical simulations show that the momentum 
transfer rate from a high Mach number jet to the ambient medium has a similar efficiency in 1D and 3D (Chernin et al., 1994).  
Therefore, it is meaningful to compare the momentum transfer rate derived from our 1D shock parameters with that measured in the CO outflow by K92.
With our estimated total shock area of 60$^{\prime\prime} \times 60^{\prime\prime}$ (from CO line intensities,
see Sec.\ref{sec:LVG}), a preshock density of n$_{\textrm{H}}$=10$^4$ cm$^{-3}$ and a shock velocity of 18 km\,s$^{-1}$, 
we infer a momentum rate of 3 $\times 10^{-4}$ M$_{\odot}$\,km\,s$^{-1}$\,yr$^{-1}$.  In comparison, K92 estimated a lower limit of 5 $\times 10^{-5}$
M$_{\odot}$\,km\,s$^{-1}$\,yr$^{-1}$ in the CO outflow, which is only a factor
of six lower than that derived by our shock model. Consequently, our proposed molecular shock is compatible with the working surface 
shock driving the whole CO outflow, although we cannot exclude the possibility that other gas components [e.g. un-shocked gas 
(Caratti o Garatti et al., 2006)] may exist and thus contribute to the momentum transfer balance.

We can further evaluate the mass-loss rate of the driving engine, if the jet speed and  the density ratio between the jet and the
ambient medium (which determines the fraction of jet momentum transferred to the cloud; e.g. Chernin et al., 1994) are known. The optical ``streamer'' south of
HH54, which is presumably tracing the ``jet'' in this system, has a typical radial velocity of -70 km s$^{-1}$ with respect to the cloud (Graham \& Hartigan,
1988), hence the jet speed is probably at least 100 km\,s$^{-1}$. Comparing the atomic line ratios in the streamer measured by the same authors with the 1D 
models of Hartigan, Morse \& Raymond (1994) indicates a jet density $\approx$ 10$^3$ cm$^{-3}$. Since our H$_2$ shock modelling suggests an ambient 
molecular cloud density of 10$^4$ cm$^{-3}$, the jet would be strongly under-dense, and the efficiency of momentum transfer to the HH54 shock would be 
close to 1. The mass-loss rate in the jet would then be $<$ 3 $\times 10^{-4}$ M$_{\odot}$\,km\,s$^{-1}$\,yr$^{-1}$ / 100 km\,s$^{-1}$
= 3 $\times 10^{-6}$ M$_{\odot}$\,yr$^{-1}$.
 
\subsubsection{CO, H$_2$O and OH emission}\label{sec:LVG}

\begin{table*}
\caption{\label{tab:h2o} H$_2$O fluxes as predicted by the selected shock models.}
\begin{center}
\begin{tabular}{ccccccc}
\hline\\[-5pt]
Wavelength$^{a}$   &  Term                  &   E$_{up}$           &      C                 &      J              &   C+J             &$\tau$$^b$ \\ 
($\mu$m)     &                        &   (cm$^{-1}$)        & \multicolumn{3}{c}{(erg\,cm$^{-2}$\,s$^{-1}$)}                   &       \\ 
\hline\\[-5pt]
  75.4       & o- 3$_{21}$-2$_{12}$   &    212.16	     &1.1 10$^{-13}$   & 1.1 10$^{-13}$   & 7.2 10$^{-14}$    & $\sim$ 10      \\
  78.7*      & o- 4$_{23}$-3$_{12}$   &    300.36	     &5.1 10$^{-14}$   & 3.2 10$^{-14}$   & 1.6 10$^{-14}$    & $<<$ 1         \\
  82.0*      & o- 6$_{16}$-5$_{05}$   &    447.25	     &5.4 10$^{-14}$   & 2.9 10$^{-14}$   & 1.3 10$^{-14}$    & $<<$ 1         \\
  99.5*      & o- 5$_{05}$-4$_{14}$   &    325.35	     &1.2 10$^{-13}$   & 7.5 10$^{-14}$   & 4.2 10$^{-14}$    & $<<$ 1         \\
 125.3*      & p- 4$_{04}$-3$_{13}$   &    222.05	     &6.1 10$^{-14}$   & 4.2 10$^{-14}$   & 2.4 10$^{-14}$    & $\sim$ 1       \\
 138.5       & p- 3$_{13}$-2$_{02}$   &    142.29	     &1.3 10$^{-13}$   & 1.0 10$^{-13}$   & 8.1 10$^{-14}$    & $\sim$ 10      \\
 174.6       & o- 3$_{03}$-2$_{12}$   &    176.76	     &3.2 10$^{-13}$   & 2.4 10$^{-13}$   & 1.5 10$^{-13}$    & $\sim$ 10      \\
 179.5       & o- 2$_{12}$-1$_{01}$   &     79.50	     &6.9 10$^{-13}$   & 6.9 10$^{-13}$   & 6.9 10$^{-13}$    & $\sim$ 10      \\
 257.8*      & o- 3$_{21}$-3$_{12}$   &    212.16	     &1.9 10$^{-14}$   & 2.5 10$^{-15}$   & 1.5 10$^{-15}$    & $<<$ 1         \\
 269.3       & p- 1$_{01}$-0$_{00}$   &     37.14	     &2.4 10$^{-13}$   & 3.0 10$^{-13}$   & 2.9 10$^{-13}$    & $\sim$ 5 10$^2$\\
 398.6       & p- 2$_{11}$-2$_{02}$   &     95.17	     &2.6 10$^{-14}$   & 1.1 10$^{-14}$   & 1.4 10$^{-14}$    & $\sim$ 1 10$^2$\\
 538.3*      & o- 1$_{10}$-1$_{01}$   &     42.37	     &5.6 10$^{-14}$   & 9.6 10$^{-14}$   & 2.0 10$^{-13}$    & $\sim$ 3 10$^2$\\
\hline\\[-5pt]
\end{tabular}
\end{center}
Notes to the table: $^a$ lines whose flux changes by more than a factor of two, depending from the shock type, are marked 
with an asterisk; $^b$ maximum value of the optical depth along the profile of the (C+J) shock.
\end{table*}

Aiming to understand the diagnostic capabilities of the far-infrared lines in discriminating different shock types, 
we have considered  all three models described in Table \ref{tab:param},
even though steady-state C-type shocks have been already disregarded on the basis of the H$_2$ emission fit.
 Our 1D assumption imposes 
a major limitation on the prediction of line profiles, which can be strongly influenced 
by the geometry; hence we will limit our analysis to the computation of the brightness of the lines.
For each model, the physical parameters at each time-step along the shock have been 
iteratively used
as input of a LVG code in plane-parallel geometry. This gives as output the brightness of the lines at each temporal 
step, which have been summed up to obtain the total integrated value to be compared with the observations.

In the LVG approximation, the line ratios are constrained only by the local gas kinetic temperature and density, in addition to the ratio of the 
species number density and velocity gradient.  Hence, in this part of the analysis, we simply compare observed line ratios with those 
determined by the physical conditions provided by the shock profiles. 
The unique free parameters are the predicted absolute line brightnesses, whose ratios with the observed fluxes constrain 
for each molecular species the angular size of the emitting region, and the temperature of the dust, whose 
continuum emission has the effect of pumping the population of the low lying energy levels. The latter parameter, however, is 
relevant only for the OH emission fitting (e.g. Thai-Q-Thung et al., 1998, Offer \& van Dishoeck, 1992).

The code has been developed for the first 45 rotational levels for both ortho- and para-H$_2$O, 41 (and 61) levels for CO and 24 for OH.
Energy levels and radiative decay rates have been taken from the LAMDA database (Leiden Atomic and Molecular Database)\footnote{available 
at http://www.strw.leidenuniv.nl/~moldata}. In the H$_2$O model, collisions with p-H$_2$ and o-H$_2$ are considered
separately in the temperature range 5-140 K (Dubernet \& Grosjean, 2002; Grojean, Dubernet \& Ceccarelli, 2003, Phillips, Maluendes \& Green, 1996), 
while at higher temperatures (up to 2000 K) the collisional rates with He have been 
rescaled to obtain a first order approximation for collisions with H$_2$ (Green, Maluendes \& McLean, 1993). 
We have assumed an ortho-to-para H$_2$O ratio of 3, as the large uncertainties associated 
with the observations do not allow us to discriminate among different values of this parameter.
In the case of the OH model we have
taken the collisional coefficients with o-H$_2$ and p-H$_2$ given by Offer, van Hernert \& van Dishoeck (1994) in the range
from 15 to 300 K. Finally, for the CO molecule, we have explored the sensitivity of the model to the 
different collisional rates available in the literature. McKee et al. (1982) give the 
$\gamma$$_{J0}$ coefficients, scaled from the collisional rates with He, to compute the downward rates for levels with 
J$_{up}$$\le$60 and 200 K $\le$ T $\le$ 2000 K: this dataset
seems therefore appropriate for line predictions in high-temperature regime. The low-temperature regime, on the contrary, 
has been investigated by Flower (2001) who has computed the collisional rates in the range 5 to 400 K, including energy levels
up to J$_{up}$=29 and J$_{up}$=20 for collisions with para-H$_2$ and ortho-H$_2$, respectively. Extrapolations to include energy levels up to
J$_{up}$=40 and collisional temperatures up to 2000 K have been made by Sch\"{o}ier et al. (2005).  

The results for the CO line predictions are plotted in Fig.\ref{fig:CO}: the long-dashed line refers to 
the collisional rates from McKee (1982), while the short-dashed one to the rates from Flower (2001) and Sch\"{o}ier et al. (2005).
Major differences (up to one order of magnitude in the predicted line brightness) occur for J$_{up}$ $\la$ 10, i.e. for
levels mainly populated at the low temperatures. This 
difference is particularly evident in the J-type shock model (middle panel): adopting the coefficients of McKee et al. (1982), we obtain a 
two-component distribution, with peaks at J$_{up}$$\simeq$8 and J$_{up}$$\simeq$16, while the coefficients of Flower (2001) and Sch\"{o}ier et al. (2005) 
produce a distribution of fluxes with a single peak at J$_{up}$$\simeq$16. Both sets of coefficients give similar results in the range of the ISO observations, 
which are rather well fitted by both the steady-state C-type shock and by the (C+J) shock.  On the other hand, the predictions of the steady-state J-type model 
deviate significantly from the observations. Therefore, we conclude (having already disregarded the C-type shock on the basis of the 
H$_2$ fit) that the (C+J) model is the only one able to account for both H$_2$ and CO observations, a result underlying the 
importance of a multi-species analysis in modelling line emission from  shocks.\\

The angular size found from the ratio of the theoretical line brightness, from the (C+J) model, of the 16-15 line (at 162 $\mu$m) with the 
observed flux, is about 40$^{\prime\prime}$, regardless of the adopted set of coefficients. 
By considering that the LWS aperture (80$^{\prime\prime}$) represents the Airy disk 
at $\approx$100\,$\mu$m of a point-like source (Swinyard et al., 1996), it follows that the derived dimension is a lower limit 
to the true size of the emitting region. This latter, 
taking into account the diffraction effects at 162 $\mu$m, can be estimated to be approximately 
60$^{\prime\prime}$, i.e. a slightly larger region than that traced by the H$_2$ pure rotational emission.  
The model indicates that the far-infrared lines arise mainly from gas at T$\approx$ 500-1000 K, which is in agreement with the temperature found 
from consideration of the \h \, 0-0 lines in the Boltzmann diagram. Transitions with 6$\la$ J$_{up}$ $\la$10 (at wavelengths 
between 260 and 433 $\mu$m) are predicted to be from a few to an order of magnitude brighter than in the other two models: their observation 
would thus provide a stringent test to the (C+J) model. This will be possible in the near future with 
the HIFI spectrometer aboard the Herschel satellite, but also from the ground with the new sub-mm APEX 
telescope (up to the $J$=8-7 transition). At present, a map of the CO 1-0 emission has been reported by K92, from which a 
flux of $\approx$ 4 $\times 10^{-16}$ erg\,s$^{-1}$\,cm$^{-2}$  is derived by integrating over $\approx$ 1 arcmin$^2$.
Noticeably, this value is only a factor of 2.5 lower than that predicted by the C+J model, if the coefficients of 
Flower (2001) and Sch\"{o}ier et al. (2005) are adopted.

\begin{figure}
\centering
\includegraphics[width=9cm]{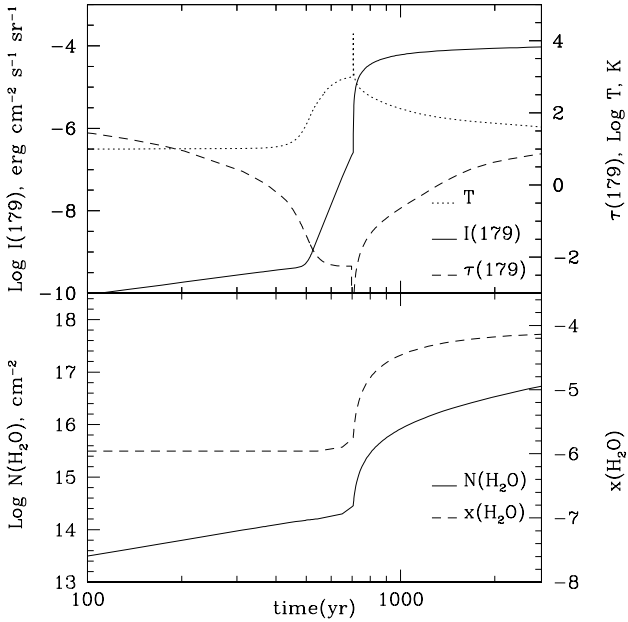}
\caption{\label{fig:179} {\it Top:} Output of the LVG model (in the framework of the C+J model): 179\,$\mu$m line brightness and optical depth.
The neutral temperature is also shown for comparison. {\it Bottom:}  Water column density and abundance predicted by the (C+J) model.}
\end{figure}

Fig. \ref{fig:h2O} reports the fits to the H$_2$O ortho and para lines. All three shock models provide good fits to the observed lines, with the marginal 
exception of the steady-state C-type shock model: the poor diagnostic capability of the LWS water lines is due 
to their similar excitation temperatures (of $\approx$ 50-100 K). 
To search for lines more sensitive to the shock parameters, we have reported in the figure the line fluxes predicted for bright H$_2$O lines 
falling in the spectral range covered by the two Herschel spectrometers PACS and HIFI.  Indeed, under the physical conditions determined by the considered 
shock models, some fluxes of lines (mainly at the shortest wavelengths) are predicted 
to change by more than a factor of two. These, indicated by an asterisk in Tab. \ref{tab:h2o}, will be all observable 
with a signal-to-noise ratio larger than ten with an integration time of few minutes with PACS and 
less than one hour with HIFI (Poglitsch, Waelkens \& Geis, 2001, de Graauw \& Helmich, 2001). In the same table, we also report the maximum value of the
line optical depth along the (C+J) shock profile: we note that lines originating in low-lying levels (those with E$_{up}$ $\la$ 200 K) are optically thick, being
the largest $\tau$ values found in correspondence of backbone transitions. We also show in Fig. \ref{fig:179} (top panel) the brightness and optical depth of the 179\,$\mu$m 
line predicted by the (C+J) model. The line brightness increases to its final value of $\sim$ 10$^{-4}$ erg cm$^{-2}$ s$^{-1}$ sr$^{-1}$ 
with the sudden increase of the temperature at the J-shock front; a corresponding decrease in the optical depth is seen at the same point. 
In the bottom panel, the water column density and abundance are plotted:
the latter increases at the shock front because of the activation of very efficient  high-temperature reactions (at T$\ga$ 300 K), which convert all oxygen not locked in CO into water
(e.g. Bergin, Neufeld, \& Melnick, 1998). In the post-shock region,  N(H$_2$O) levels out at $\approx$ 6 10$^{16}$ cm$^{-2}$. At the same time, the average
abundance $<$x(H$_2$O)$>$ over the far-IR emitting region, determined as the ratio N(H$_2$O)/N(H$_2$) integrated over the shock up to 3 10$^3$ yr, is $\sim$ 7 10$^{-5}$.
This roughly agrees with the estimates by Liseau et al. (1996) and also with the values found in a number of 
outflows from Class 0 sources (Giannini, Nisini \& Lorenzetti, 2001). Thus, the discrepancy between water abundance measurements and predictions from 
classical non-dissociative shock models (x(H$_2$O) larger than 10$^{-4}$, Kaufman \& Neufeld, 1996) can be better reconciled in the framework of J-type shock with magnetic precursor
models.  Moreover, since gaseous water survives only in high-temperature regions,
we expect that its emission arises in compact parts of the shock: by normalizing the predicted 179\,$\mu$m line brightness to the 
observed flux, we derive, even considering the diffraction at the line wavelength, $\theta$ $\approx$30 arcsec. Therefore, the 
water emission region is more compact than that emitting the warm CO by about a factor of four.  Noticeably, a difference
in the shock area for CO and H$_2$O lines, signals some inconsistencies in 1D geometry, since in this approximation the gradients within 
the cooling zone (perpendicular to the shock surface) are already taken into account and thus the emitting area should
be exactly the same for all the species.

As regards the OH molecule, we have detected only two lines at 119 and 163 $\mu$m, 
each connecting the lowest energy levels of the two ladders ($^2\Pi_{3/2}$ and $^2\Pi_{1/2}$) into which the OH energy level diagram is split.
These lines have a similar excitation temperature and critical density, thus not allowing us from  
discriminating between different shock categories. Using the C+J shock structure, the LVG computation predicts a 119/163 ratio of 14, 
which is substantially different from the observed value of 2. This discrepancy could be due to the pumping effect of a dust continuum, the 
heating of which is not accounted for in our shock model.  Indeed, the 163\,$\mu$m line brightness is increased more than that of the 119 $\mu$m 
by dust pumping (Offer \& van Dishoeck, 1992).
 
In summary, our multi-species analysis leads us to consider the (C+J) shock as the model which best reproduces the overall observed molecular 
emission. The parameters relative to the FIR emission are given in  Tab. \ref{tab:fitparam}. Here, we report the molecular column densities and the fractional
abundances predicted by the (C+J) shock model and integrated along the shock profile, along with the angular sizes of the emitting regions derived 
by the ratio of the predicted brightnesses with the observed fluxes.

\begin{table}
\caption{\label{tab:fitparam}Molecular parameters derived for the J-type shock with magnetic precursor model.}
\begin{center}
\begin{tabular}{ccccc}
\hline\\[-5pt]
Species             & N                & $\theta$	&   $<$x$>$           \\
                    & (cm$^{-2}$)      &  (arcsec)	& 	        \\
\hline\\[-5pt]
H$_2$               &7.4 10$^{20}$     &       -        &    -          \\
CO                  &1.2 10$^{17}$     &       60       &  1.6 10$^{-4}$  \\
H$_2$O              &5.4 10$^{16}$     &       30       &  7.3 10$^{-5}$\\
OH                  &1.3 10$^{14}$     &       -        &  1.8 10$^{-7}$ \\
\hline\\[-5pt] 
\end{tabular}
\end{center}
\end{table}

\subsubsection{Atomic emission}\label{sec:ion}

\begin{table}
\caption{\label{tab:OI} [{\oi}]63,145$\mu$m brightness as predicted by  shock models.}
\begin{center}
\begin{tabular}{ccc}
\hline\\[-5pt]
Model  & I(63)            &     I(145)\\
       &\multicolumn{2}{c}{(erg\,cm$^{-2}$\,s$^{-1}$ sr$^{-1}$)}\\ 
\hline\\[-5pt]       
 C     &  9.9  10$^{-7}$  &    9.1 10$^{-8}$\\
 J     &  3.2  10$^{-8}$  &    1.5 10$^{-9}$\\  
 C+J   &  1.3  10$^{-6}$  &    1.4 10$^{-7}$\\    
\hline\\[-5pt]
\end{tabular}
\end{center}
\end{table}

Of the atomic and ionic species detected in HH54, oxygen is the only one involved in the same chemical reactions
as H$_2$O and OH.  Therefore, it seems reasonable to expect that a sizeable fraction of the emission at 63 and 145 $\mu$m should arise in 
the same shock as the molecular lines. In Table \ref{tab:OI} we give the line brightness of the 
[{\oi}] 63 and 145 $\mu$m predicted by the three shock models
considered in this paper. Even for the highest values (predicted by the (C+J) model), an emitting area fifty times larger
than the ISO beam is required in order to reproduce the observed flux of the 63 $\mu$m line. This suggests that oxygen emission originates 
in gas excited by a different mechanism to that which gives rise to the  H$_2$ line emission. Both the 
FWHM of the 63\,$\mu$m line profile and the observed 63/145 ratio of 23 are typical of radiative J-shock environments (e.g. Hollenbach \& McKee, 1989),
which could form at the apex of the bow-shock or in reverse shocks along the jet. If we 
compute the mass-flux from the 63\,$\mu$m flux using the formula  \.M$_{shock}$ = 10$^{-4}$ M$_{\odot}$/yr$\times$ (L(63
$\mu$m)/L$_{\odot}$) (e.g. Liseau et al., 1997) we find \.M$_{shock}$ = 1.2 10$^{-6}$ M$_{\odot}$/yr, which is similar to the jet mass flux 
estimated in Sec.\ref{env}.

With regard to the other atomic and ionic species, none of the observed lines can be accounted for by the (C+J) shock model: the predicted 
intensities of [\fe], [\ci] and [\s] are approximately 4 orders of magnitude smaller than are observed. Indeed, Spitzer maps (Neufeld et
al. 2006) have shown that the bulk of the fine structure emission (from [{\ion{Ne}{ii}}], [{\ion{S}{i}}],[{\ion{Si}{ii}}],[{\ion{Fe}{ii}}]) 
comes from a defined region close to the head of the bow: 
our results suggest that the degree of dissociation and ionization in this region are much higher than in the gas in which the H$_2$ emission originates.
The same authors state that the fine structure lines 
originate in fast J-type shocks of velocity $\approx$ 35-90 km\,s$^{-1}$, in agreement with the (rough) estimate we derived from the  
[\oi]\,63\,$\mu$m line for the shock velocity. \\
In conclusion, the atomic/ionic emission appears to trace a different (faster and more ionizing) shock than the H$_2$ emission. Similar results have already 
obtained by us when fitting the near-infrared lines observed in HH99 (M$^{\mathrm{c}}$Coey et al., 2004): in that case, in a self-consistent scenario, 
two J-shock components, one with a magnetic precursor and one fully dissociative, were required to account for the H$_2$ and the ionic emissions, 
respectively. However, we postpone a full analysis of this topic to a later paper.

\section{Conclusions}
We have presented a multi--frequency/multi--species analysis of the molecular emission in HH54, by means of
one of the most complete spectroscopic data-base in the infrared (from 1 to 200 $\mu$m) ever collected
for an HH object. 
The H$_2$ lines (1-12 $\mu$m) coming from levels $v$=0 to $v$=4 with excitation energies between 2000 and 25000 K have been 
interpreted in the context of a state-of-art shock code. Two models (a steady-state J-type shock and a quasi-steady J-type shock 
with magnetic precursor) yield satisfactory fits to the data, allowing us to discard steady-state C-type shocks as a mechanism for the line excitation.
The output parameters (temperature and density profiles, molecular species number density and velocity gradient) from these models are adopted as 
input to an LVG computation to interpret the emission of CO, H$_2$O and OH in the far-infrared. This second step of the analysis allows us to definitively 
select the J-type shock with magnetic precursor  as the one able both to account for the overall emission of H$_2$, CO and H$_2$O.
None of the considered models, however, are able to reproduce the observed OH emission, which probably could be accounted
for by considering the heating of dust grains in the shock model. A further, fully dissociative, 
high-velocity shock is thought to be responsible for the observed atomic/ionic lines. If it is the case, a picture may emerge
in which the optical and near-IR ionic lines trace a fragmented jet shock, the near-IR H$_2$ lines trace ambient J-shocks driven by the jet shocks, and the
$v$=0 H$_2$ lines trace the continuos precursor propagating into the ambient cloud.\\
Finally, the adopted approach has been extended to provide predictions of the best diagnostic H$_2$O line intensities in the
range accessible to the forthcoming facilities HIFI and PACS aboard the Herschel satellite.
   
\vspace{0.5cm}
\emph{Acknowledgements}: 
We thank the referee for a detailed and constructive report.
S.C. gratefully acknowledges the support of Marc Sauvage with ISOCAM data planning and analysis tools.
This work was partially supported by the European
Community's Marie Curie Research and Training Network JETSET (Jet
Simulations, Experiments and Theory) under contract MRTN-CT-2004-005592.

\end{document}